\documentclass[journal]{IEEEtran}

\usepackage[utf8]{inputenc}
\usepackage{cite}
\usepackage[usenames,dvipsnames]{xcolor}
\usepackage[pdftex]{graphicx}
\usepackage{circuitikz}

\usepackage[hidelinks,bookmarks=false]{hyperref}
\usepackage{booktabs,colortbl}
\usepackage{multirow}
\usepackage[cmex10]{amsmath}
\usepackage{mathrsfs}
\usepackage{mathtools}

\DeclarePairedDelimiter\floor{\lfloor}{\rfloor}

\usepackage{array}

\usepackage{siunitx}
\usepackage{amssymb}

\newcommand{\eg}{{e.g.\ }}
\newcommand{\citefig}[1]{Fig.~\ref{#1}}

\newcommand{\citesec}[1]{{Section~\ref{#1}}}

\newcolumntype{L}[1]{>{\raggedright\let\newline\\\arraybackslash\hspace{0pt}}m{#1}}
\newcolumntype{C}[1]{>{\centering\let\newline\\\arraybackslash\hspace{0pt}}m{#1}}
\newcolumntype{R}[1]{>{\raggedleft\let\newline\\\arraybackslash\hspace{0pt}}m{#1}}

\usepackage{xstring}
\newcommand{\paper}[1]{{{\ignorespaces
  Paper~\textbf{\IfEqCase{#1}{%
    {pes}{[\ref{paper:pes}]}%
    {isgt}{[\ref{paper:isgt}]}%
    {freezer}{[\ref{paper:freezer}]}%
    {repl}{[\ref{paper:repl}]}%
    {upec}{[\ref{paper:upec}]}%
    {cdc}{[\ref{paper:cdc}]}%
    {gm}{[\ref{paper:gm}]}}[\PackageError{paper}{Paper not defined: #1}{}]%
}}}}

\usepackage{tikz}
\usetikzlibrary{decorations.pathreplacing}
\usetikzlibrary{arrows,shapes,positioning}
\usetikzlibrary{patterns}
\usepackage{pgfplots}
\pgfplotsset{compat=newest}
\pgfplotsset{plot coordinates/math parser=false}
\newlength\figureheight
\newlength\matlabfigurewidth
\ifCLASSOPTIONcompsoc
 \usepackage[caption=false,font=normalsize,labelfont=sf,textfont=sf]{subfig}
\else
 \usepackage[caption=false,font=footnotesize]{subfig}
\fi

\usepackage{multirow}
\usepackage{booktabs,colortbl}

\begin{document}

%
\title{Achieving the Dispatchability of Distribution Feeders through Prosumers Data Driven Forecasting and Model Predictive Control of Electrochemical Storage}

\author{%
\IEEEauthorblockN{Fabrizio~Sossan,~\IEEEmembership{Member,~IEEE}, Emil~Namor,~\IEEEmembership{Student~Member,~IEEE}, Rachid Cherkaoui, ~\IEEEmembership{Senior Member,~IEEE},  Mario~Paolone,~\IEEEmembership{Senior Member,~IEEE}.}

\thanks{The authors are with the Distributed Electrical Systems Laboratory, École Polytechnique Fédérale de Lausanne, Switzerland (EPFL), e-mail: \{fabrizio.sossan, emil.namor, rachid.cherkaoui, mario.paolone\}@epfl.ch.

This research received funding from the Swiss Competence Center for Energy Research (FURIES) and Swiss Vaud Canton within the initiative ``100 millions pour les énergies renouvelables et l'efficacité énergétique''.
}
}

\markboth{IEEE Trans. On Sustainable Energy, to be published (2016).}%
{}

\maketitle

\begin{abstract}
We propose and experimentally validate a process to dispatch the operation of a distribution feeder with heterogeneous prosumers according to a trajectory with 5 minutes resolution, called \emph{dispatch plan}, established the day before the operation. The controllable element is a utility-scale grid-connected battery energy storage system (BESS) integrated with a minimally pervasive monitoring infrastructure. The process consists of two stages: day-ahead, where the \emph{dispatch plan} is determined by using forecast of the aggregated consumption and local distributed generation (prosumption), and real-time operation, where the mismatch between the actual prosumption realization and dispatch plan is compensated for thanks to adjusting the real power injections of the BESS with model predictive control (MPC). MPC accounts for BESS operational constraints thanks to reduced order dynamic grey-box models identified from on-line measurements.  
The experimental validation is performed by using a grid-connected 720~kVA/500~kWh BESS to dispatch the operation of a 20~kV distribution feeder of the EPFL campus with both conventional consumption and distributed photo-voltaic generation.
 
\end{abstract}


\section{Introduction}
\IEEEPARstart{T}{he} progressive displacement of conventional generation in favor of renewables requires restoring an adequate capacity of regulating power to assure reliable operation of interconnected power systems. 
An emerging concept to tackle this challenge is achieving the controllability of portions of distribution networks by exploiting controllable distributed generation (DG), flexible demand and storage. 
This paradigm can be traced in a number of frameworks, such as virtual power plants (VPPs) and grid-tied microgrids, which, in broad terms, consist in operating aggregates of heterogeneous resources with the objective of providing ancillary services to the upper grid layer, \eg\ \cite{biegel2014aggregation, 5558756, Soshinskaya2014659}. In general, solutions based on aggregating the capability of DERs require extended ICT infrastructures and efficient control policies to harvest flexibility until the LV distribution level, see \eg \cite{Bernstein2015, costanzo2013coordination, 7038106}.
As a matter of fact, these solutions are of difficult integration in the existing grid at the current stage because: (i) they might not offer the same reliability level as conventional generation, (ii) they require a radical change of the operational practices, and (iii) their technical requirements are not met or expensive to implement.

An essential aspect to enable the transition towards a \emph{smarter grid} is the availability of plug-and-play technologies that can provide grid ancillary services in the current operational and regulatory framework with a reduced set of technical requirements and minimal levels of complexity. An example of research efforts in this direction is given by the proposition of directly coupling storage with stochastic renewable DG in order to achieve dispatchable operation, as e.g. in \cite{teleke2009control, marinelli2014testing, 6913566}. 

In this paper, we merge the two aforementioned concepts and propose to achieve the dispatchability of distribution feeders with DG by controlling utility-scale battery energy storage systems (BESSs). 
Specifically, we describe a process to dispatch the operation of a group of prosumers (characterized by both conventional demand and photo-voltaic, PV, generation) according to a scheduled prosumption trajectory at 5 minutes resolution, called \emph{dispatch plan}, established the day before operation by implementing forecast of the local prosumption\footnote{We therefore do not perform any intra-day redispatch.}. During real-time operation, the mismatch between the \emph{dispatch plan} and prosumption realization is corrected by adjusting the real power injections of the BESS converter with model predictive control (MPC). MPC is designed to track the dispatch plan while obeying to BESS voltage, current and state-of-charge (SOC) constraints, which are modeled thanks to prediction models. 
Both battery and consumption forecasting models are data-driven (namely identified from experimental measurements).


The proposed process is experimentally validated by dispatching the operation of a 20~kV medium voltage active distribution system of the EPFL campus, called \emph{dispatchable feeder}, by using a grid-connected Lithium Titanate 720~kVA/500~kWh BESS placed at the root of the feeder. The \emph{dispatchable feeder} includes office buildings with conventional demand and 95~kWp rooftop PV DG. It relies on a minimally invasive monitoring infrastructure and requires only the measurements of the power flow at the grid connection point (GCP) and the information from the battery management system (BMS).

The envisaged benefits of achieving dispatched operation by design of distribution systems are two-fold. At the system level, it allows to reduce the uncertainty associated with prosumption operation, which is known to increase balancing power requirements and system costs, see e.g. \cite{5590013, lu2013nv}. At the local level, the \emph{dispatch plan} is generated in order to meet the requirements of local distribution systems operation, for example, to perform consumption peak-shaving. 

In comparison with methodologies that have been proposed in the literature to achieve the controllability of heterogeneous clusters of prosumers, the proposed process is characterized by a low complexity in terms of metering, computation and communication requirements. Indeed, the principal coordination mechanism is given by the commitment of the operator to track the dispatch plan, without involving complex iterations or real-time communication with an upper level aggregator or coordinator. 
Moreover, the two-stage formulation similar to the conventional way of operating the power system (day-ahead planning and intra-day operation) could facilitate the integration in the current grid operational paradigm. 

The contributions of this paper are:
\begin{itemize}
 \item formulation of a BESS-based control strategy to dispatch the operation of a group of heterogeneous stochastic and uncontrollable prosumers according to a trajectory established the day before the beginning of the operation;
 \item a short-term energy management MPC strategy for the BESS which implements predictive constraints on the BESS current, voltage and SOC while retaining a convex formulation of the underlying optimization problem;
 \item experimental validation of the proposed method on a real-scale and real-life grid.
\end{itemize}

The paper is organized as follows: \citesec{sec:statement} states the problem and, to fix ideas, introduces the experimental setup used for the validation. Sections \ref{sec:dayahead} and \ref{sec:realtime} respectively describe day-ahead and real-time operation. Section~\ref{sec:results} presents the experimental results, and, finally, Section~\ref{sec:concs} concludes the paper by summarizing the contributions.

\section{Preliminaries and Problem Statement}\label{sec:statement}
We consider a distribution network populated by an unknown mix of electric loads, possibly with distributed generation too, and equipped with a BESS. 
The considered scenario is well exemplified by the adopted experimental configuration, which is depicted in \citefig{fig:networksetup}. It includes a group of buildings of the EPFL campus with 350~kW peak consumption (equipped with 95~kWp root-top PV installations) and a grid-connected 720~kVA/500~kWh Lithium Titanate BESS. 
The BESS bidirectional real power flow is denoted by $B$, while $P$ is the composite power flow as seen at the GCP.
The former is the control variable\footnote{The BESS is equipped with a four quadrant converter. However, only the real power axis is considered in this application.}, and its actual realization is available from measurements. The latter is known by information from a phasor measurement unit (PMU) placed at the root of the feeder \cite{EPFL-CONF-203775}. The aggregated feeder demand is denoted by $L$, and, by neglecting grid losses\footnote{The targeted grid has a radial topology and characterized by co-axial cables line with a cross section of \SI{95}{\square\milli\meter} and a length of few hundreds meters. Therefore, the grid losses are negligible. See \cite{EPFL-CONF-203775} for further details.}, it is estimated as $L=P-B$. In general, $L$ consists of both demand and generation and, in the following, it will be simply denoted as \emph{prosumption}. For both buildings and BESS, we use the passive sign notation, namely positive values denote consumption, while negative denote injections towards the GCP. At the current stage, we assume that the local distribution feeder has  ample capacity and characteristics always to operate within its voltage and lines ampacities technical constraints. 

\begin{figure}[!ht]
\centering
\scalebox{0.85}
{
{\footnotesize
\input{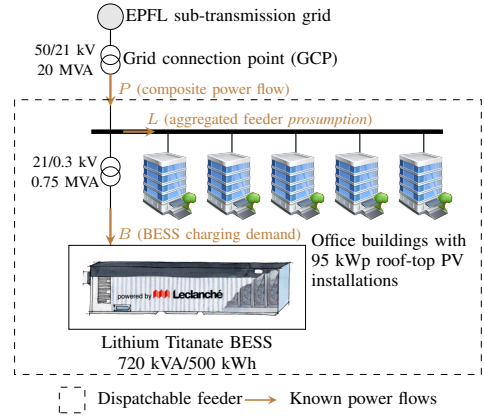}
}
}
\caption{The experimental setup used to validate the dispatchable feeder concept: a MV feeder with a number of buildings of the EPFL campus equipped with roof-top PV installations, and a BESS. The control framework relies on a minimally invasive monitoring infrastructure: it requires the knowledge of the composite power flow at the GCP and the BESS injection.} \label{fig:networksetup}
\end{figure}

The problem consists in dispatching the feeder such that the composite power transit at the GCP follows a power consumption sequence with 5 minutes resolution, called \emph{dispatch plan}, which is established the day before the beginning of the operation. Similarly to the conventional way of planning power system operation, the control strategies consists in a two-stage process:

\begin{itemize}
 
 \item {\bf Day-ahead operation}. The dispatchable feeder operator determines the \emph{dispatch plan} based on forecast on the prosumption and accounting also for the energy necessary to restore an adequate BESS state-of-energy (SOE) to establish a minimum level of up/down regulation capacity. The \emph{dispatch plan} is a sequence of average power consumption values with 5 minutes resolution that the feeder should follow during the next day of operation. The day-ahead procedure is repeated every day. At the current stage, we arbitrarily choose to perform it one hour before the beginning of the real-time operation. At clock-time 00:00~UTC of the next day, the \emph{dispatch plan} comes into effect and real-time operation begins. This phase is detailed in \citesec{sec:dayahead}.
 
 \item {\bf Real-time operation}. This phase starts at clock-time 00:00~UTC of the day of operation, and lasts for the next 24~hours period.  The dispatchable feeder operator adjusts the BESS power injection $B$ to compensate for the mismatch between the \emph{dispatch plan} and the power prosumption realization, which are likely to occur due to forecasting errors. This is accomplished by applying model predictive control (MPC), accounting for the dynamic behavior of the BESS voltage and SOC thanks to dynamic grey-box models identified from BESS measurements, and including short-term forecasts of the prosumption. This procedure is detailed in \citesec{sec:realtime}.
\end{itemize}

\noindent The operation sequence is also sketched in the timeline of \citefig{fig:comm}, which shows the interactions between the load balance responsible (as for example described in \cite{balancegroup}), the dispatchable feeder operator (which implement the proposed control strategy) and the BESS.

The choice of the 5 minute dispatch interval is according to the envisaged trend for real-time electricity markets. Although not specifically discussed in this work, this formulation potentially allows for day-ahead scheduling considering also dynamic electricity prices, in a similar way as done in \cite{6009220, 6740918, 7308089}.

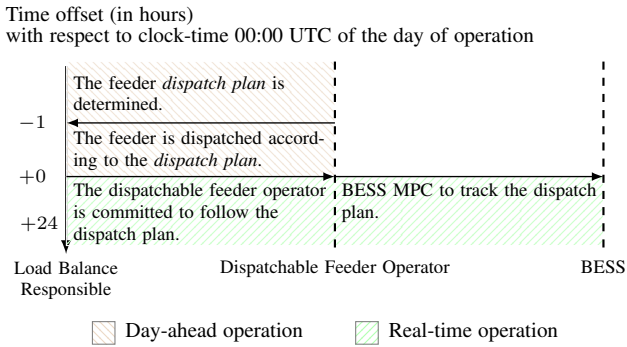
\begin{figure}[!h]
\centering
\scalebox{1}{
{\scriptsize
\begin{tikzpicture}
\tikzset{>=latex}
\tikzstyle{int}=[fill=white, minimum size=0.5cm, draw=gray!80, thick]

\begin{scope}[shift={(-2.35, 0)}]

\draw[thick, ->] (0,3) node[text width=9cm, xshift=3.7cm, yshift=15]{{\footnotesize Time offset (in hours)\\ with respect to clock-time 00:00~UTC of the day of operation}} -- (0,0.5) node[below, text width=2.2cm, text centered]{\scriptsize Load Balance Responsible};
\end{scope}

\begin{scope}[shift={(-2.3, 0)}]

\scalebox{1.02}{

\draw[pattern=north west lines, pattern color=brown!20!white, draw=white] (0,3) rectangle (3.5,1.5);
\draw[pattern=north east lines, pattern color=green!20!white, draw=white] (0,1.5) rectangle (7,0.60);

\draw[thick, dashed] (3.5,3) -- (3.5,0.5) node[below, text width=5.2cm, text centered]{\scriptsize Dispatchable Feeder Operator};
\draw[thick, dashed] (7,3) -- (7,0.5) node[below, text width=2.2cm, text centered]{\scriptsize BESS};

\draw[line width=0.5, <-] (0,2.2) node[above right, text width=3.45cm, yshift=2]{The feeder \emph{dispatch plan} is determined.} node[left]{$-1$~~~} -- (3.5,2.2) ;
\draw[line width=0.5, ->] (0,1.5) node[left](zeroam){$+0$~~~} node[above right, text width=3.5cm]{The feeder is dispatched according to the \emph{dispatch plan}.} node[below right, text width=3.45cm]{The dispatchable feeder operator is committed to follow the dispatch plan.}-- (3.5,1.5) ;
\node[below=0.25cm of zeroam]{$+24$};
\draw[line width=0.5, ->] (3.5,1.5) node[below right, text width=3.45cm]{BESS MPC to track the dispatch plan.} -- (7,1.5) ;
}
\end{scope}

\begin{scope}[shift={(-2, -0.7)}]
\draw[pattern=north west lines, pattern color=brown!30!white, draw=gray] (0, 0) rectangle (0.3, 0.3);
\node[text height=1.5ex, text width=2.5cm, anchor=west] at (0.35, 0.15){\footnotesize Day-ahead operation};
\end{scope}

\begin{scope}[shift={(1.5, -0.7)}]
\draw[pattern=north east lines, pattern color=green!30!white, draw=gray] (0,0) rectangle (0.3, 0.3);
\node[text height=1.5ex, text width=3.5cm, anchor=west] at (0.35, 0.15){\footnotesize Real-time operation};
\end{scope}

\end{tikzpicture}
}}
\caption{Dispatchable feeder operation: the day before operation, the dispatchable feeder operator determines the \emph{dispatch plan} for the next day.} \label{fig:comm}
\end{figure}

\section{Day-ahead problem}\label{sec:dayahead}
The objective is to determine the \emph{dispatch plan}, which is the power consumption profile at 5 minutes resolution that the feeder should follow during the next day operation. The \emph{dispatch plan} is denoted by the sequence $\widehat{P}_i,\ i=0,1,\dots,N-1$ where the index $i$ represents the 5-minute interval of the day of operation, and $N=288$ is the number of 5-minute intervals in 24~hours. It is defined as:
\begin{align}
 \widehat{P}_i = \widehat{L}_i + F^o_i, && i=0,\dots,N-1,\label{eq:dispatchplan}
\end{align}
where $\widehat{L}_0, \dots, \widehat{L}_{N-1}$  are prosumption point predictions, determined as described in \ref{sec:ppredictions}, and the sequence $F^o_0, \dots, F^o_{N-1}$ is called offset profile (the notation $^o$ means ``optimal'' and denotes the output of an optimization problem, as detailed later in this section). The latter accounts for the amount of BESS energy necessary to restore an adequate level of flexibility. For example, at the end of each day of operation, the BESS residual charge might be close to the upper or lower bounds; the offset plan biases the \emph{dispatch plan} such that the BESS will discharge or charge, therefore re-establishing a suitable energy level to compensate for the mismatch between the \emph{dispatch plan} and the prosumption realization. As similarly envisaged in \cite{6913566}, scheduling the BESS energy demand in the \emph{dispatch plan} to manage its residual charge intrinsically  enables continuous-time operation. Especially, it avoids the need of assuming an initial state of energy (SOE) level or imposing a dedicated constraint on the final SOE at the end of the day of operation (see for example \cite{marinelli2013testing, teng2013optimal, 5545425}).

Albeit several methods may be alternatively used to calculate the offset plan and implement this intuitive concept, we here decide to adopt a framework based on robust convex optimization (see Section~\ref{sec:offsetplan}). The main reason for this choice is that convex optimization is a well-established discipline for which there exist several mature off-the-shelf software libraries, which guarantees a robust and bug-free implementation while allowing to focus only on the problem formulation.

\subsection{Day-ahead Prosumption Forecasting} \label{sec:ppredictions}
Although being a well established methodology for high levels of aggregation, consumption and generation forecasting at local level is a relatively unexplored topic. It recently come to prominence especially in connection to large scale deployment of distributed generation in distribution networks, microgrids operation, flexible demand management and thanks to the progressive availability of metering infrastructure at the low-voltage level, e.g. \cite{6417004,  bacher2013, 7232358, Sossan20161}. Local prosumption is characterized by a high volatility due to the reduced spatial smoothing effect, in case for example of PV generation, and prominence of isolated stochastic events, such as induction motors inrushes due to the insertion of pumps or elevators. 
Because of these reasons, existing forecasting methodologies developed considering high levels of aggregation, e.g. \cite{1350819}, might fail in forecasting low populated aggregates of prosumers.

For the proposed application, we apply a fairly simple non-parametric forecasting strategy that is able to capture the prosumption daily and seasonal components with a decent level of accuracy. As it will be detailed in the following, it consists in selecting and averaging a number of historical power consumption sequences that are relevant for the period to predict according to the value of certain prosumption data features.
It is important to note that the described dispatch process is independent of the selected forecasting method. In particular, in this work, we do not claim specific contributions in the context of prosumption forecasting, and a more exhaustive assessment of the performance of the proposed forecasting tool is postponed to future works.

\subsubsection{Training data set}
We consider a series of historical power prosumption measurements collected at the GCP at 5~minutes resolution organized in daily sequences of length $N$ (the bold typeface denotes a sequence of scalars obtained by stacking the time evolving value of the referenced variable):
\begin{align}
 \boldsymbol{l}_{y, d} = \left[ l_{y,d,0},l_{y,d,1},\dots, l_{y,d,N-1} \right] \label{eq:historicalpowerconsumption}
\end{align}
where $y$ and $d$ respectively denote the calendar year ($2015, 2016, \dots$) and calendar day-of-year ($1, 2, \dots, 365$) of the observation. The observations at 5~minutes resolution are obtained by average downsampling the original historical time series, which is at 20~milliseconds resolution. Also, historical measurements of the daily global horizontal radiation (\SI{}{\kilo\watt\hour\per day\per\square\meter}) are available and are denoted by $R_{y,d}$: each data point is obtained by integrating observations at 10~minutes resolution of the global horizontal irradiance (GHI) over a 24~hours period (source MeteoSwiss, meteostation ``Lausanne freiland'', GPS coordinates \ang{06;38;56}\! \ang{46;33;33}\!).

\subsubsection{Forecast computation}
Assuming being on a given day, the objective is to determine the prosumption forecast at 5~minutes resolution for the time interval from hour 0 UTC (Coordinated Universal Time) to 24 UTC of the incoming day, called \emph{target~day}. The information associated with the target~day are the calendar day-of-year $d^*$, calendar year $y^*$ and the forecasted global horizontal radiation $R^*$. The last information is obtained by applying the model in \cite{26540804} starting from forecast of the cloud coverage (obtained for the considered experimental location by using NorwayMet services). The cloud coverage to GHI conversion model requires the clear-sky radiation, which is computed for Lausanne by using the clear-sky model implementation available in the open source geographical information system GRASS \cite{hofierka2002solar, neteler2012grass}, which allows accounting for the topological shading induced by geographical features on the horizon.

The first step to compute the prosumption forecast is to select from the historical data a number of sequences $\boldsymbol{l}_{y, d}$ with similar features as those of the target day. This is done by applying the following heuristic:

\begin{itemize}
 
 \item a set $\Omega_A$ is composed by selecting among historical observations the couples $(y,d)$ which refer to calendar non-working days (weekends and bank holidays) if $(y^*,d^*)$ correspond to a non-working day, and vice-versa.
 
 \item A set $\Omega_B$ is obtained by selecting the first 10 couples $(y, d)$ in $\Omega_A$ chosen  in increasing order of time-distance with respect to the target~day. The time-distance is evaluated as $365\cdot (y-y^*) + \left|d-d^* \right|$.
 
 \item A set $\Omega_C$ is obtained by selecting the first 5 couples $(y, d)$ in $\Omega_B$ chosen in increasing order of distance between the respective observed cumulated irradiance $R_{y,d}$ and the target radiance $R^*$, evaluated as $\left| R^* - R_{y,d} \right|$.
\end{itemize}
Summarizing, the set $\Omega_C$ is composed of 5 couples of indexes $(y,d)$ which are i) closest in time to the target-day, ii) same kind as the target day, and iii) closest in amount of radiation to the GHI forecast for the target-day.

Finally, the couples in $\Omega_C$ are used to select as many sequences of historical prosumption measurements \eqref{eq:historicalpowerconsumption}, which are regarded to as potential realizations of the next day prosumption. 
In particular, the set with the estimates of the prosumption realization is given as:
\begin{align}
 \mathscr{L}_i = \left\{ l_{d,y,i}\ \forall (d,y) \in \Omega_C \right\},
\end{align}
for each time interval $i=0,1,2,\dots,N-1$. At the current stage, realizations from the uncertainty sets are assumed independent in time, in other words, time correlation is not modeled.The prosumption point predictions $\widehat{L}_0, \dots, \widehat{L}_{N-1}$ are obtained by averaging the elements in the respective estimates set:
\begin{align}
 \widehat{L}_i = \dfrac{1}{\left|\mathscr{L}_i\right|} \sum_{l \in \mathscr{L}_i} l, && i=0, 1, \dots, N-1,
\end{align}
where $\left|\mathscr{L}_i\right|$ denotes the number of elements in $\mathscr{L}_i$ (set cardinality).
It is to mention that the outdoor temperature is not included as an influential variable in the selection process of the historical profiles: this is because the considered group of buildings are not equipped with electric space heating. Moreover, the dependency between PV generation and temperature is neglected at this stage.

\subsection{Determination of the Offset Profile} \label{sec:offsetplan}
The BESS injection required during real-time operation to compensate the mismatch between the dispatch plan and the prosumption realization $L_i$ is given as:
\begin{align}
 B_i &= \widehat{P}_i - L_i = \widehat{L}_i + F^o_i - L_i,  && i=0,\dots,N-1 \label{eq:dayahead:binjection0} 
\end{align}
where \eqref{eq:dispatchplan} has been used in the last equality to explicitly express the BESS power ${B}_i$ as a function of the consumption forecast uncertainty and offset plan. Since $L_i$ is unknown at the current time, we model it as a random realization $l_i$ from the uncertainty set $\mathscr{L}_i$. Therefore from \eqref{eq:dayahead:binjection0}, an estimate $\widehat{B}_i$ of the BESS compensation action at time interval $i$ is:
\begin{align}
 \widehat{B}_i =  F^o_i + \widehat{L}_i - l_i, && l_i \in \mathscr{L}_i. \label{eq:dayahead:binjection}
\end{align}

The smallest and largest BESS power realization are respectively:
\begin{align}
\inf{\left\{\widehat{B}_i\right\}} = F^o_i + \widehat{L}_i - \underset{l_i \in \mathscr{L}_i }{\sup}{\{l_i\}} = F^o_i + L^\downarrow_i \label{eq:infB} \\
\sup{\left\{\widehat{B}_i\right\}} = F^o_i + \widehat{L}_i - \underset{l_i \in \mathscr{L}_i }{\inf}{\{l_i\}} = F^o_i + L^\uparrow_i \label{eq:supB}
\end{align}
where the quantities $L^\downarrow_i = \widehat{L}_i - l_i^\uparrow$ (negative by construction) and $L^\uparrow_i = \widehat{L}_i - l_i^\downarrow$ (positive by construction) are introduced.

In this part of the problem, we use the notion of BESS SOE, the evolution of which is expressed as the following linear function of the BESS power:
\begin{align}
 \text{SOE}_{i+1} = \text{SOE}_{i} + \beta(\widehat{B}_i) \cdot \widehat{B}_i \label{eq:dayahead:soc}
\end{align}
where $\beta(\widehat{B}_i)$ is the charge/discharge efficiency (we neglect at this stage its dependency on temperature and power magnitude)
\begin{align}
 \beta(\widehat{B}_i) =
 \begin{cases}
 \beta^+ = T_s/3600 \cdot \eta,  &  \widehat{B}_i \ge 0 \\
 \beta^- = T_s/3600 \cdot {1}/{\eta}, & \widehat{B}_i < 0,
 \end{cases}\label{eq:conversion:efficiency}
\end{align}
$T_s=300$~s (5 minutes) is the length of the discretization step, and $\eta$ is the conversion efficiency, assumed constant\footnote{ We note that $\eta$ is normally a function of the charge/discharge rate. However, we neglect this dependency at this stage.} and estimated from measurements considering values of the BESS charge/discharge rate close to the operating ones (i.e. C/5 in our case). We re-write \eqref{eq:dayahead:soc} as:
\begin{align}
 \text{SOE}_{i+1} = \text{SOE}_{i} + \beta^+ \left[\widehat{B}_i\right]^+ - \beta^+ \left[\widehat{B}_i\right]^-
\end{align}
where the operator $[\cdot]^+$ is
\begin{align}
 [x]^+ = \begin{cases}
        x, & x > 0 \\
        0, & \text{otherwise}
       \end{cases}
\end{align}
and viceversa for $[\cdot]^-$.

The lowest and highest possible BESS SOE are given by propagating for $i=0,\dots,N-1$ the dynamic equation \eqref{eq:dayahead:soc} from a known initial state-of-charge $\text{SOE}^\downarrow_{0} = \text{SOE}^\uparrow_{0} = \text{SOE}_0$ (\footnote{Since the dispatch plan is computed one hour before the beginning of the day of operation, the initial BESS state-of-charge $\text{SOE}_0$ is unknown and should be therefore predicted. At the current stage we use a persistent predictor, namely $\text{SOE}_0 = \text{SOC}_{-5\times 12} $.}) accounting for the smallest and largest $\widehat{B}_i$ in \eqref{eq:infB} and \eqref{eq:supB}, respectively:
\begin{align}
& \text{SOE}^\downarrow_{i+1} = \text{SOE}^\downarrow_{i} + \beta^+ \left[ F^o_i + L^\downarrow_i \right]^+ + \beta^- \left[ F^o_i + L^\downarrow_i \right]^- \label{eq:soedown:nonlinear}\\
& \text{SOE}^\uparrow_{i+1} = \text{SOE}^\uparrow_{i} + \beta^+ \left[ F^o_i + L^\uparrow_i \right]^+ + \beta^- \left[ F^o_i + L^\uparrow_i \right]^- \label{eq:soeup:nonlinear}
\end{align}
The two expressions above are embedded into the following optimization problem, which has the objective of determining the offset plan $\boldsymbol{F}^o=[{F}^o_0, \dots,{F}^o_{N-1}]$ such that BESS SOE and power are always in the respective allowed bounds $(\text{SOE}_\text{min}, \text{SOE}_\text{max})$ and $(B_\text{min}, B_\text{max})$ in the worse case scenarios:

\begin{align}
 \boldsymbol{F}^o = \underset{ 
 \renewcommand*{\arraystretch}{0.8}
 \begin{matrix}
  \scriptstyle \boldsymbol{F} \in \mathbb{R}^{N}
 \end{matrix}}
 {\text{arg~min}}\left\{ \sum_{i=1}^{N} F^2_i \right\}\label{eq:dayahead:cost}
\end{align}
subject to
\begin{align}
 & \text{SOE}^\downarrow_{i+1} = \text{SOE}^\downarrow_{i} + \beta^+ \left[ F^o_i + L^\downarrow_i \right]^+ + \beta^- \left[ F^o_i + L^\downarrow_i \right]^- \label{eq:dayahead:icm1} \\
 & \text{SOE}^\uparrow_{i+1} = \text{SOE}^\uparrow_{i} + \beta^+ \left[ F^o_i + L^\uparrow_i \right]^+ + \beta^- \left[ F^o_i + L^\uparrow_i \right]^- \label{eq:dayahead:icm0} \\
 & \text{SOE}^\downarrow_{i+1} \ge \text{SOE}_\text{min} , &&  \label{eq:dayahead:ic0}\\
 & \text{SOE}^\uparrow_{i+1} \le \text{SOE}_\text{max} \label{eq:dayahead:ic1} \\
 & F_i + L^\downarrow_i \ge B_\text{min}  \label{eq:dayahead:ic2}\\
 & F_i + L^\uparrow_i \le B_\text{max} \label{eq:dayahead:ic3}\\
 & \widehat{P}_i \le P_\text{max} \label{eq:dayahead:ic4},
\end{align}
for $i=0,\dots,N-1$, $\text{SOE}^\downarrow_{0} = \text{SOE}^\uparrow_{0} = \text{SOE}_0$ and  $\text{SOE}_0$ is given.
The inequality constraint \eqref{eq:dayahead:ic4} is to make sure that the dispatch plan is smaller than a tunable threshold, the value of which can be designed by the modeller to peak shave the real power power transit at the GCP (as it will be shown in the Results Section).
The formulation in \eqref{eq:dayahead:cost}-\eqref{eq:dayahead:ic4} is nonconvex due to the relationship \eqref{eq:dayahead:icm1}-\eqref{eq:dayahead:icm0}. In the real implementation and experiments, we solve an equivalent convex formulation of it, which is shown in Appendix \ref{appendix:convexformulation}.

It is important to note that the analysis proposed in this paper is limited to the formulation and validation of the dispatch plan and of the associated control problem. In particular, aspects related to the BESS sizing problem are beyond the scope of this paper and will be covered in future works. 

\subsection{Implementation of the day-ahead strategy}
Summarizing, the day-ahead strategy consists in first computing the forecast for the next day of operation $\widehat{L}_0,\dots,\widehat{L}_{N-1}$, as described in \ref{sec:ppredictions}. Therefore, the offset profile $F^o_0, \dots, F^o_{N-1}$ is computed by solving the optimization problem \eqref{eq:dayahead:cost}-\eqref{eq:dayahead:ic4}, finally allowing the computation of the dispatch plan $\widehat{P}_0,\dots,\widehat{P}_{N-1}$ with \eqref{eq:dispatchplan}.
The day-ahead strategy is implemented as a Matlab script and it is scheduled for operation every day at hour 23 UTC. The implementation is on a Intel i5 PC with Debian OS.

%
\section{Real-time operation}\label{sec:realtime}
The objective of the real-time operation is to adjust the BESS real power injection such that the average power consumption at the end of each 5-minute period matches the respective set-point from the dispatch plan. The control objective is formalized in Paragraph \ref{sec:cobj:formulation}. Beforehand, we introduce the following notation:

\begin{itemize}
 \item the control strategy is actuated with a sample time of 10~s; the period has been chosen in order to capture early time BESS dynamics and assure good control performance.
 
 \item The index $k=0,1,2,\dots,K-1$ denotes the rolling 10~seconds time interval of the current day of operation, where $K=8640$ is the number of 10~seconds period in 24~hours.
 
 \item At the beginning of each interval $k$, the real power flow at the GCP, the BESS flow, and the real power of the prosumption realization for the previous interval $k-1$ become known thanks to measurements. They are respectively denoted by $P_{k-1}$, $B_{k-1}$ and $L_{k-1}$. We note that, although the BESS power flow is our control variable, its actual realization might differ due to imprecise actuations of the converter and BESS transformer losses.

 \item The value of the prosumption set-point to match, denoted by $P^*_k$, is retrieved from the \emph{dispatch plan} $\widehat{P}_0, \widehat{P}_1, \dots, \widehat{P}_{N-1}$ as:
 \begin{align}
P^*_k = \widehat{P}_{ \floor*{ \frac{k}{30} }  } \label{eq:rt:setpoint}, 
\end{align}
where $\floor*{ \cdot }$ denotes the nearest lower integer of the argument, and $30$ is the number of 10-second intervals in a 5-minute slot.

\item The $k$-index of the first 10-second interval for the current 5-minute slot is denoted as $\underline{k}$ and is:
\begin{align}
 \underline{k} = \floor*{ \frac{k}{30} } \cdot 30.
\end{align}
For example, at clock-time 00:16~UTC (sixteen minutes past midnight), $k=96$, $P^*_k=\widehat{P}_3$, and $\underline{k}=90$. Similarly, the $k$-index of the last 10-second interval for the current 5-minute slot is:
\begin{align}
 \overline{k} = \underline{k} + 30 - 1 \label{eq:rt:uplimit}.
\end{align}

 \item The control action consists in actuating the set-point for the BESS converter real power demand (in kilowatt, kW). It is denoted by $B^o_k$ and is piecewise constant in the interval $k$. It is determined by model predictive control as described in Paragraph~\ref{sec:mpc}.
 
 \end{itemize}

\begin{figure}[!ht]
\centering
\begin{tikzpicture}
\newcommand{\myfactor}{1.5}


\begin{scope}[shift={(0.3,0)}]
\draw [decorate,decoration={brace,amplitude=3pt},xshift=-0pt,yshift=0pt]
  (\myfactor*0.5*0,0.5) -- (\myfactor*0.5*1,0.5) node [black,midway,yshift=10pt]{{\small $L_{0}$}};
\draw [decorate,decoration={brace,amplitude=3pt},xshift=-0pt,yshift=0pt]
  (\myfactor*0.5*1,0.5) -- (\myfactor*0.5*2,0.5) node [black,midway,yshift=10pt]{{\small $L_{1}$}};
  
\node at (\myfactor*0.5*5.0,0.84) {{\scriptsize (Power consumption measurements)}};

\draw [decorate,decoration={brace,amplitude=3pt},xshift=0pt,yshift=0pt]
  (\myfactor*0.5*3,-0.1) -- (\myfactor*0.5*2,-0.1) node [text width=2.5cm, align=center, midway, yshift=-12pt]{{\scriptsize $B^o_{2}$\par}};

\foreach \x in {0,...,8} {
  \draw [line width=0.4mm, color=gray] (\x*\myfactor*0.5, 0.05) -- (\x*\myfactor*0.5, -0.05) node[anchor=north] {};
}

\draw[->] (0,0) -- (4.0*\myfactor+0.5,0) node[anchor=north, text width=2cm, text centered] {\footnotesize{}};
\draw [line width=0.5mm] (0*\myfactor, 0.1) -- (0, -0.5) node[anchor=north] {};
\draw [line width=0.5mm] (3.5*\myfactor, 0.1) -- (3.5*\myfactor, -0.5) node[anchor=north] {};

\draw [decorate,decoration={brace,amplitude=3pt},xshift=-0pt,yshift=0pt]
  (3.5*\myfactor,-0.75) -- (0*\myfactor,-0.75) node [black,midway,yshift=-10pt]{\small{$\widehat{P}_0$ {\scriptsize (from \emph{dispatch plan}) }}};
  
\draw [decorate,decoration={brace,amplitude=3pt},xshift=-0pt,yshift=0pt]
  (4.15*\myfactor,-0.75) -- (3.5*\myfactor,-0.75) node [black,midway,yshift=-10pt]{\small{$\widehat{P}_1$}};


\foreach[count=\x] \l in {0, 1,2,3,4,$\dots$,29,30,31,\hphantom{11}Index $k$} {
  \draw (\x*\myfactor*0.5-\myfactor*0.5, 0.1) node[anchor=south] {\scriptsize{\l} };
}

\begin{scope}
\draw [decorate,decoration={brace,amplitude=3pt},xshift=0pt,yshift=0pt]
  (\myfactor*0.5*1,-0.1) -- (\myfactor*0.5*0,-0.1) node [text width=2.5cm, align=center, midway, yshift=-12pt]{{\small $B^o_{0}$\par}};
\draw [decorate,decoration={brace,amplitude=3pt},xshift=0pt,yshift=0pt]
  (\myfactor*0.5*2,-0.1) -- (\myfactor*0.5*1,-0.1) node [text width=2.5cm, align=center, midway, yshift=-12pt]{{\small $B^o_{1}$\par}};
  \node at (\myfactor*0.5*4.6,-0.50) {{\scriptsize (Battery set-points)}};
\end{scope}

\end{scope}
\end{tikzpicture}
\caption{The first 31 10-second intervals of the day of operation. It is sketched the situation at the beginning of the time interval 2.} \label{fig:timeline}
\end{figure}
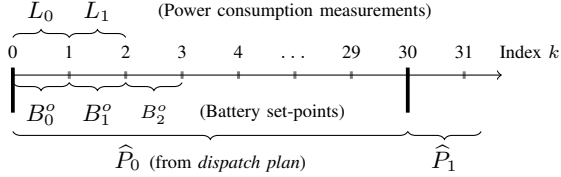

\noindent The nomenclature is also exemplified in Fig.~\ref{fig:timeline}, which sketches, for example, the situation at the beginning of the time interval $k=2$ (clock-time 00:00:20~UTC): the BESS real power set-points $B^o_0$ and $B^o_1$ were actuated already in the previous two intervals, $B^o_2$ has been just determined using the most recent information (namely, the prosumption realizations $L_0$ and $L_1$), and the average prosumption set-point to achieve in the 5 minute interval is given by the first value of the dispatch plan, $\widehat{P}_0$.


\subsection{Formulation of the Control Objective}\label{sec:cobj:formulation}
Say being at the beginning of the time interval $k$, the average composite power flow at the GCP (prosumption + BESS injection) for the current 5 minute slot is given by averaging the available information until $k$. If $k$ corresponds to the beginning of a 5 minute period, no information is available yet, and we say that the average composite consumption is zero. Formally, it is:
\begin{align}
 P_{k} = 
 \begin{cases}
0 & k = \underline{k}\\
 \dfrac{1}{k-\underline{k}} \cdot \sum\limits_{j=\underline{k}}^{k-1} \left(L_j + B_j \right) & k > \underline{k}.
 \end{cases}\label{eq:Pavg}
\end{align}
For example: at clock-time 00:05~UTC (300~seconds since the beginning of the day, $k=\underline{k}=20$), the average power consumption is zero; twenty~seconds later ($k=22, \underline{k}=20$), two samples are available and averaged to determine $\overline{P}_{22}$. %

We augment the definition in \eqref{eq:Pavg} by including short-term point predictions  of the prosumption $\widehat{L}_{k|k},\widehat{L}_{k+1|k}, \dots, \widehat{L}_{\overline{k}|k}$. This allows to to calculate the \emph{expected} average composite consumption for the whole duration of the current 5 minute slot. By exploiting the definition given in  \eqref{eq:Pavg}, the expected average consumption accounting for the short-term point predictions is:
\begin{align}
  P^+_{k} = \dfrac{1}{30}\left( (k-\underline{k})\cdot P_k + \sum\limits_{j=k}^{\overline{k}} \widehat{L}_{j|k}\right).\label{eq:Ppavg}
\end{align}
At the current stage, we use a persistant predictor\footnote{The use of more advanced short-term prediction models will be considered in future works.} to model future realizations, namely $\widehat{L}_{j|k}=L_{k-1}, \ j=k,\dots,\overline{k}$.

The energy error, expressed in kWh, between the realizations and the dispatch plan in the current 5-minute slot (\emph{dispatch plan error}):
\begin{align}
e_{k} =  \dfrac{300}{3600} \cdot \left(P^*_k  - P^+_{k}\right), \label{eq:rt:dpe}
\end{align}
where 300 and 3600 are respectively the number of seconds in a 5 minutes interval and 1 hour interval, $P^*_k$ and $P^+_{k}$ are as defined in \eqref{eq:rt:setpoint} and \eqref{eq:Ppavg}, respectively. 



\subsection{Formulation of Model Predictive Control }\label{sec:mpc}
In general, MPC consists in determining the control action for a given system by solving at each time step an optimization problem with updated information, where the system constraints are enforced by implementing prediction models in the optimization problem. MPC has been widely adopted in several engineering fields, and proposed also in application to power system operation, for example for control of demand response and storage \cite{5530680, Hredzak201584}.
In this work, MPC is to determine the BESS real power injection $B^o_k$ in order to achieve zero \emph{dispatch plan error} \eqref{eq:rt:dpe} by the end of each 5-minute slot while obeying to BESS SOC, DC voltage and current operational limits.

The main reason for implementing the control problem with a MPC framework rather than a PID-based regulator is the following: for the latter, the control action depends only on current and past values; MPC allows to schedule the whole power trajectory withing the targeted time horizon, therefore, it is more suitable to solve an energy management problem. For example, assuming that reliable short-term predictions indicate that there will be a significant mismatch between realization and dispatch plan in the second half of a certain 5-minute time slot, the predictive control framework could preemptively react while respecting BESS operational constraints thanks to enforcing them explicitly in the formulation. On the contrary, a feedback control loop reacts when measurements are available and might have larger chance to fail if the error is large or too close to the end of the 5-minute interval.


%

In the MPC optimization problem, there are two potential candidates for decision variables: \emph{i)}, the real power injection on the AC side, and \emph{ii)}, the current on the DC side. We adopt the second solution because it admits a convex equivalent formulation of the optimization problem, as shown in the following. We explicitly seek for convexity because convex optimization problems, besides having a single minimum point which is the global (provided that the solution exists), can be solved efficiently and in a reliable way.


\paragraph{BESS Energy Throughput}
We model the BESS energy throughput (in \si{\kilo\watt\hour}) on the AC bus in the discretized time period from $k$ to $\overline{k}$:
\begin{align}
 E_{\overline{k}|k}(v_k, \dots, v_{\overline{k}}, i_k, \dots, i_{\overline{k}}) = \alpha \sum_{j=k}^{\overline{k}} v_j i_j, \label{eq:energytp}
\end{align}
where $v_k$ and $i_k$ are the battery DC voltage and current (positive when charging and vice-versa), respectively, and the scale factor $\alpha=10/3600 \cdot 0.98$ is to convert from power (in kW) in the discretized 10~seconds time interval to energy (in kWh), and the coefficient 0.98 models the converter losses (estimated from measurements). The formula above can be expressed as the following matrix product:
\begin{align}
 E_{\overline{k}|k}(\cdot) = \alpha \boldsymbol{v}_{\overline{k}|k}^T\boldsymbol{i}_{\overline{k}|k}\label{eq:EN|tpre},
\end{align}
where the bold notation denotes sequences obtained by stacking in column vectors the realizations in time of the referenced variables, e.g. $\boldsymbol{v}_{\overline{k}|k} = \begin{bmatrix}v_k, \dots, v_{\overline{k}} \end{bmatrix}^T$. Since the voltage on the DC bus is modeled by using a linear electrical circuit (as detailed in \ref{sec:predictions:voltage}), its dynamic evolution can be expressed as a linear function of the battery current. By applying the transition matrices $\phi^v, \psi^v_i, \psi^v_1 $ (which are developed starting from the voltage discrete state-space model representation as described in Section~\ref{sec:predictions:voltage} and Appendix~\ref{sec:MPCtransitionmatrices}), the battery voltage can be expressed by:
\begin{align}
 \boldsymbol{v}_{\overline{k}|k} = \phi^v x_k + \psi^v_i \boldsymbol{i}_{\overline{k}|k} + \psi^v_1 \boldsymbol{1}\label{eq:voltagedynamic},
\end{align}
where $x_k$ is the state vector of the voltage model and is known from measurements. 

It is to note that the voltage evolution also depends on the BESS SOC. As it will be detailed in Section~\ref{sec:predictions:voltage}, this dependency is captured by identifying five linear models for as many different SOC ranges. The linear transition matrices in \eqref{eq:voltagedynamic} are obtained by selecting the appropriate voltage model according to the current measured BESS SOC (\emph{model scheduling}). Moreover, we assume that the BESS SOC does not vary significantly in the actuation period, so we can write the voltage as a time invariant linear function of the BESS current, therefore retaining linearity.

Replacing \eqref{eq:voltagedynamic} into \eqref{eq:EN|tpre} yields to: 
\begin{align}
 \begin{aligned}
 E_{\overline{k}|k}(\boldsymbol{i}_{\overline{k}|k}) = \alpha \left(\phi^v x_k + \psi^v_i \boldsymbol{i}_{\overline{k}|k} + \psi^v_1 \boldsymbol{1}\right)^T \boldsymbol{i}_{\overline{k}|k} = \\
 = \alpha \left({x}^T_k {\phi^v}^T \boldsymbol{i}_{\overline{k}|k} + \boldsymbol{i}^T_{\overline{k}|k} {\psi^v_i}^T \boldsymbol{i}_{\overline{k}|k} + \boldsymbol{1}^T {\psi^v_1}^T \boldsymbol{i}_{\overline{k}|k}\right),
 \end{aligned}
 \label{eq:EN|t}
\end{align}
where $\boldsymbol{1}$ denotes the all-ones vector.

We now explore the requirements for $E_{\overline{k}|k}(\cdot)$ in \eqref{eq:EN|t} being a convex function of $\boldsymbol{i}_{\overline{k}|k}$. Since the nonnegative sum between functions preserves convexity, the convexity of \eqref{eq:EN|t} requires that all its three addends are convex. The first and third addend are linear in $\boldsymbol{i}_{\overline{k}|k}$, therefore convex. The second term is a quadratic form of $\boldsymbol{i}_{\overline{k}|k}$: the necessary and sufficient condition for its convexity is given by $\psi^v_i$ being semidefinite positive. This hypothesis has been verified numerically for all the BESS identified voltage models, which are presented in Section~\ref{sec:predictions:voltage}. We conclude that $E_{\overline{k}|k}$ in \eqref{eq:EN|t} is convex in $\boldsymbol{i}_{\overline{k}|k}$.

\paragraph{Formulation and Implementation}
The energy tracking problem (given by achieving zero \emph{dispatch plan error} at the end of the 5-minute slot) could be formulated by minimizing the squared deviation of \eqref{eq:EN|t} from \eqref{eq:rt:dpe}, such as:
\begin{align}
\left(E_{\overline{k}|k}(\boldsymbol{i}_{\overline{k}|k}) - e_k \right)^2.
\end{align}
However, as known from the functions composition rules \cite{boyd_convexoptimization}, the convexity of $p(x)=q(r(x))$ when $r(x)$ is convex requires $q$ convex non decreasing, which is not this case because the squared function of the difference is convex but not nondecreasing on all its domain. Therefore, we reformulate the objective and achieve a convex equivalent formulation of the original problem: it consists in maximizing the BESS DC current while imposing that the convex BESS energy throughput \eqref{eq:EN|t} is smaller than or equal to the energy target \eqref{eq:rt:dpe}; this achieves the energy throughput to hit the upper bound of the inequality\footnote{In order this equivalent formulation to hold, the BESS energy throughput must be a monotonically increasing function of the current. Since we apply a linear battery equivalent circuit where parameters are assumed constant in the 10~seconds actuation period, the power output (thus its integral, the energy) monotonically increases with the current until the value imposed by the maximum power transfer theorem (which, for our case, is in the range of 1000~A, well above the operational current limit of the considered BESS). The same can be also verified numerically by taking the first derivative of \eqref{eq:EN|t} with respect to $\boldsymbol{i}_{\overline{k}|k}$ and observing that all components are positive for the typical operating range of BESS current.}, thus achieving the same value as the target energy. The combination of a linear cost function with an inequality constraint in the form ``$f(x) \le 0$, with $f$ convex in $x$'' is a convex optimization problem.

Overall, the formulation of the MPC optimization problem is given by augmenting the just described formulation with \emph{i)} constraints on the BESS current and its rate of change and \emph{ii)} open open-loop predictive constraints on BESS voltage and BESS state-of-charge (SOC). It is worth noting that, whereas in the day-ahead problem the constraints on the BESS flexibility were enforced by considering the BESS power and SOE, in this case we consider BESS current, SOC and voltage. This allows for a higher degree of modeling detailing and is indeed more suitable for the control problem, where the primary objective is determining a control decision that is respectful of BESS operation constraints.

Formally, the decision problem is:
\begin{align}
 \boldsymbol{i}^o_{\overline{k}|k} = \underset{\boldsymbol{i} \in \mathbb{R}^{(k-\overline{k}+1)}}{\text{arg~max}} \left\{ \boldsymbol{1}^T \boldsymbol{i}_{\overline{k}|k} \right\} \label{eq:rt:mpc0}
\end{align}
subject to
\begin{align}
 & \alpha\left({x}^T_k {\phi^v}^T \boldsymbol{i}_{\overline{k}|k} + \boldsymbol{i}^T_{N|t} {\psi^v_i}^T \boldsymbol{i}_{\overline{k}|k} + \boldsymbol{1}^T {\psi^v_r}^T \boldsymbol{i}_{\overline{k}|k}\right) \le e_k & \label{eq:rt:mpc1} \\
 & \boldsymbol{1}\cdot{i}_\text{min}  \preccurlyeq \boldsymbol{i}_{\overline{k}|k} \preccurlyeq \boldsymbol{1}\cdot{i}_\text{max} \label{eq:rt:mpc2}\\
 & \boldsymbol{1}\cdot \Delta_{i,\text{min}} \preccurlyeq H \boldsymbol{i}_{\overline{k}|k} \preccurlyeq \boldsymbol{1}\cdot \Delta_{i,\text{max}} \label{eq:rt:mpc22} \\
 & \boldsymbol{v}_{\overline{k}|k} = \phi^v {v}_k + \psi^v_i \boldsymbol{i}_{\overline{k}|k} + \psi^v_1 \boldsymbol{1}  \label{eq:rt:mpc3} \\
 & \boldsymbol{1}\cdot{v}_\text{min} \preccurlyeq \boldsymbol{v}_{\overline{k}|k} \preccurlyeq \boldsymbol{1}\cdot{v}_\text{max} \label{eq:rt:mpc4} \\
 & \text{\bf{SOC}}_{\overline{k}|k} = \phi^\text{SOC} \text{SOC}_k + \psi^\text{SOC}_i \boldsymbol{i}_{\overline{k}|k} \label{eq:rt:mpc5} \\ 
 & \boldsymbol{1}\cdot \text{SOC}_\text{min} \preccurlyeq \text{\bf{SOC}}_{\overline{k}|k} \preccurlyeq \boldsymbol{1}\cdot \text{SOC}_\text{max} \label{eq:rt:mpc6},
\end{align}
where $\boldsymbol{i}^o_{\overline{k}|k} \in \mathbb{R}^{(k-\overline{k}+1)}$ is the computed control action trajectory, $\boldsymbol{1}$ denotes the all-ones column vector, the multiplication $\boldsymbol{1} \cdot \gamma$ denotes the all-$\gamma$ column vector, and the symbol $\preccurlyeq$ is the component-wise inequality. The cost function \eqref{eq:rt:mpc0} consists in maximizing the sum of the equally weighted current values over the shrinking horizon from $k$ to $\overline{k}$. This, in combination with the inequality \eqref{eq:rt:mpc1}, achieves the BESS energy throughput to be as close as possible to $e_k$, as introduced earlier. The inequalities \eqref{eq:rt:mpc2} and \eqref{eq:rt:mpc22} respectively enforce minimum and maximum magnitude and rate of change for the BESS current, where $(i_\text{min}, i_\text{max})$ and $(\Delta_{i,\text{min}}, \Delta_{i,\text{max}})$ are the respective limits and the matrix $H \in \mathbb{R}^{(k-\overline{k}+1)\times(k-\overline{k}+1)}$ is:
\begin{align}
 H=\begin{bmatrix}1 & 1 & 0 & 0 & \dots & 0 \\ 0 & 1 & 1 & 0 & \dots & 0 \\ \vdots & \vdots & \vdots & \vdots & \ddots & \vdots \\ 0 & 0 & 0 & \dots & 1 & 1  \end{bmatrix}.
\end{align}
The equality \eqref{eq:rt:mpc3} is the electrical equivalent circuit model of the the BESS according to the notation previously discussed for \eqref{eq:voltagedynamic}, while \eqref{eq:rt:mpc4} imposes BESS voltage limits, which are denoted by the couple $\left({v}_\text{min}, {v}_\text{max}\right)$.

Analogously, the equality constraint in \eqref{eq:rt:mpc4} is the evolution of the BESS SOC as a linear function of the variable $\boldsymbol{i}_{\overline{k}|k}$, where $\phi^\text{SOC}, \psi^\text{SOC}_i$ are transition matrices calculated as shown in Appendix~\ref{sec:MPCtransitionmatrices} using the SOC model described in \ref{eq:pmodel:soc}. Finally, \eqref{eq:rt:mpc6} imposes the limits on the BESS SOC, which are given by the couple $\left(\text{SOC}_\text{min}, \text{SOC}_\text{max}\right)$.

The optimization problem \eqref{eq:rt:mpc0}-\eqref{eq:rt:mpc6} is convex since the cost function is linear and all the inequality constraints are convex in $\boldsymbol{i}_{\overline{k}|k}$. 
It is noteworthy that if the real power injection had instead being adopted as the optimization variable, the problem would not have been convex because the BESS voltage evolution is nonlinear in the power and thus the constraints in \eqref{eq:voltagedynamic} would have been nonconvex.


The optimization problem is solved at each time step $k$ (with updated information) on a shrinking horizon from the index $k$ to $\overline{k}$, namely from current time until the end of the current 5-minute slot. At each $k$, the control trajectory for the whole residual horizon is available, however only the first component of the current control law is considered for actuation, which we denote by $i^o_k$.
Since the BESS power flow is controlled by using a real power reference signal, it is required to transform from $i^o_k$ to the power set-point $B^o_{k}$. By using the same model applied in \eqref{eq:energytp}, it is:
\begin{align}
 B^o_{k} = v_k \cdot i^o_k \label{eq:powersetpoint}.
\end{align}

Since the control decision is re-evaluated every 10~seconds, errors on the voltage predictions and short-term consumption forecast which arise in the current actuation period are absorbed in the next cycle, where updated measurements are used. A summary and a diagram to summarize the real-time procedure and MPC operation is proposed at the end of this section, in \ref{sec:realtimeimplementation}.

It is noteworthy that the current implementation does not prevent the tracking problem to fail, for example if the accumulated mismatch between realizations and \emph{dispatch plan} is larger than the BESS capacity.
The discussion concerning mechanisms to manage real-time contingencies and proper storage capacity sizing is postponed to future investigations.

\subsection{Prediction Models} \label{sec:predictions}

\subsubsection{BESS Voltage} \label{sec:predictions:voltage}
\paragraph{Model Formulation}
Battery voltage models for control application are normally based on electric equivalent circuits, which trade detailed modelling of the electrochemical reactions for increased tractability, see for example \cite{1634598, YannLiaw2004835, bahramipanahenhanced}. A voltage model commonly adopted in the literature is the so-called two time constants model (TTC): it consists in a linear second order dynamic model, where the value of the model parameters depend on the battery SOC, temperature and C-rate.
In this work, we augment the classical TTC formulation by implementing grey-box modelling, a set of rigorous and systematic methods for data-driven dynamic model identification (see for example \cite{Sossan20161, 4336b714bb4b4ad8869e62d0bf115bf5}). It consists in increasing the complexity of the model (by adding a order, for example) until reaching a satisfactory behavior of the model prediction errors, which at the final stage should be i.i.d. (independent and identically distributed) random noise with zero mean. This is with the main objective of obtaining a prediction model that is tuned to capture accurately the dynamics of the BESS that is used in the experimental validation of the control framework. The adopted modelling procedure initially consists in recording a series of experimental measurements of the BESS DC voltage and current while requiring to the BESS a random power flow according to a PRBS (pseudo random binary signal). The PRBS is a two levels square wave (in this case $\pm 200~$kW) with on-off periods of random durations. It is normally adopted in model identification because it is able to excite a wide range of system dynamics. As mentioned earlier, model parameters normally depend on the BESS SOC: in order to capture this dependence, a number of PRBS experimental sessions are performed when the BESS is in different SOC ranges (0-20\%, 20-40\%, 40-60\%, 60-80\%, 80-100\%). The dependencies between model parameters and both C-rate and temperature are not modeled at this stage, and it will the objective of future investigations. Nevertheless, it is worth noting that the latter dependency is expected to play a minor role because the batteries are installed in a temperature controlled environment at \SI{20}{\celsius}.

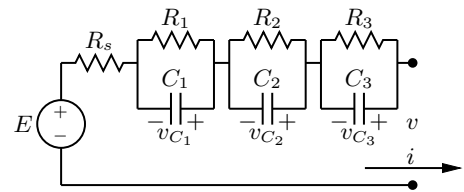
\begin{figure}[!h]
\centering
\small
\begin{tikzpicture}[scale=2.54]
\ifx\dpiclw\undefined\newdimen\dpiclw\fi
\global\def\dpicdraw{\draw[line width=\dpiclw]}
\global\def\dpicstop{;}
\dpiclw=0.8bp
\dpiclw=0.8bp
\dpiclw=0.75bp
\dpicdraw (0,0)
 --(0,0.195)\dpicstop
\dpicdraw (0,0.32) circle (0.049213in)\dpicstop
\draw (0,0.2575) node{$_-$};
\draw (0,0.3825) node{$_+$};
\dpicdraw (0,0.445)
 --(0,0.64)\dpicstop
\draw (-0.125,0.32) node[left=-1.5bp]{$ E$};
\dpicdraw (0,0.64)
 --(0.075,0.64)
 --(0.095833,0.681667)
 --(0.1375,0.598333)
 --(0.179167,0.681667)
 --(0.220833,0.598333)
 --(0.2625,0.681667)
 --(0.304167,0.598333)
 --(0.325,0.64)
 --(0.4,0.64)\dpicstop
\draw (0.2,0.681667) node[above=-1.5bp]{$ R_{s}$};
\dpicdraw (0.4,0.76)
 --(0.475,0.76)
 --(0.495833,0.801667)
 --(0.5375,0.718333)
 --(0.579167,0.801667)
 --(0.620833,0.718333)
 --(0.6625,0.801667)
 --(0.704167,0.718333)
 --(0.725,0.76)
 --(0.8,0.76)\dpicstop
\draw (0.6,0.801667) node[above=-1.5bp]{$ R_{1}$};
\dpicdraw (0.8,0.76)
 --(0.8,0.4)\dpicstop
\dpicdraw (0.8,0.4)
 --(0.625,0.4)\dpicstop
\dpicdraw (0.625,0.483333)
 --(0.625,0.316667)\dpicstop
\dpicdraw (0.575,0.483333)
 --(0.575,0.316667)\dpicstop
\dpicdraw (0.575,0.4)
 --(0.4,0.4)\dpicstop
\draw (0.6,0.483333) node[above=-1.5bp]{$ C_{1}$};
\draw (0.625,0.4) node[below right=-1.5bp]{$ +$};
\draw (0.6,0.316667) node[below=-1.5bp]{$ v_{C_1}$};
\draw (0.575,0.4) node[below left=-1.5bp]{$ -$};
\dpicdraw (0.4,0.4)
 --(0.4,0.76)\dpicstop
\dpicdraw (0.8,0.64)
 --(0.88,0.64)\dpicstop
\dpicdraw (0.88,0.76)
 --(0.955,0.76)
 --(0.975833,0.801667)
 --(1.0175,0.718333)
 --(1.059167,0.801667)
 --(1.100833,0.718333)
 --(1.1425,0.801667)
 --(1.184167,0.718333)
 --(1.205,0.76)
 --(1.28,0.76)\dpicstop
\draw (1.08,0.801667) node[above=-1.5bp]{$ R_{2}$};
\dpicdraw (1.28,0.76)
 --(1.28,0.4)\dpicstop
\dpicdraw (1.28,0.4)
 --(1.105,0.4)\dpicstop
\dpicdraw (1.105,0.483333)
 --(1.105,0.316667)\dpicstop
\dpicdraw (1.055,0.483333)
 --(1.055,0.316667)\dpicstop
\dpicdraw (1.055,0.4)
 --(0.88,0.4)\dpicstop
\draw (1.08,0.483333) node[above=-1.5bp]{$ C_{2}$};
\draw (1.105,0.4) node[below right=-1.5bp]{$ +$};
\draw (1.08,0.316667) node[below=-1.5bp]{$ v_{C_2}$};
\draw (1.055,0.4) node[below left=-1.5bp]{$ -$};
\dpicdraw (0.88,0.4)
 --(0.88,0.76)\dpicstop
\dpicdraw (1.28,0.64)
 --(1.36,0.64)\dpicstop
\dpicdraw (1.36,0.76)
 --(1.435,0.76)
 --(1.455833,0.801667)
 --(1.4975,0.718333)
 --(1.539167,0.801667)
 --(1.580833,0.718333)
 --(1.6225,0.801667)
 --(1.664167,0.718333)
 --(1.685,0.76)
 --(1.76,0.76)\dpicstop
\draw (1.56,0.801667) node[above=-1.5bp]{$ R_{3}$};
\dpicdraw (1.76,0.76)
 --(1.76,0.4)\dpicstop
\dpicdraw (1.76,0.4)
 --(1.585,0.4)\dpicstop
\dpicdraw (1.585,0.483333)
 --(1.585,0.316667)\dpicstop
\dpicdraw (1.535,0.483333)
 --(1.535,0.316667)\dpicstop
\dpicdraw (1.535,0.4)
 --(1.36,0.4)\dpicstop
\draw (1.56,0.483333) node[above=-1.5bp]{$ C_{3}$};
\draw (1.585,0.4) node[below right=-1.5bp]{$ +$};
\draw (1.56,0.316667) node[below=-1.5bp]{$ v_{C_3}$};
\draw (1.535,0.4) node[below left=-1.5bp]{$ -$};
\dpicdraw (1.36,0.4)
 --(1.36,0.76)\dpicstop
\dpicdraw (1.76,0.64)
 --(1.84,0.64)\dpicstop
\dpicdraw[fill=black](1.84,0.64) circle (0.007874in)\dpicstop
\draw (1.84,0.32) node{$v$};
\dpicdraw[fill=black](1.84,0) circle (0.007874in)\dpicstop
\dpicdraw (1.84,0)
 --(0,0)\dpicstop
\filldraw[line width=0bp](1.99625,0.065748)
 --(2.09,0.089185)
 --(1.99625,0.112623) --cycle
\dpicstop
\dpicdraw (2.068525,0.089185)
 --(1.59,0.089185)\dpicstop
\draw (1.829263,0.089185) node[above=-1.5bp]{$ i$};
\end{tikzpicture}
\caption{Equivalent circuit model of the BESS. The quantities $v$ and $i$ are respectively the BESS terminal voltage and DC current, while $v_{C_1}, v_{C_2}, v_{C_3}$ are the components of the state vector $x$ in \eqref{eq:ssd:1}-\eqref{eq:ssd:2}.}\label{fig:modelstructure}
\end{figure}

The BESS voltage model is formulated by using the continuous time stochastic state-space model representation:
\begin{align}
& dx = \mathcal{A}_c({\theta}){x} dt + \mathcal{B}_c({\theta}) {u(t)} dt + \mathcal{K}_c ({\theta}) d{\omega} \label{eq:ssd:1} \\
& v_{k} = \mathcal{C} {x}_{k} + \mathcal{D}({\theta}) {u}_{k} + \mathcal{G}({\theta}) g_k\label{eq:ssd:2},
\end{align}
where ${x} \in \mathbb{R}^{n}$ is the system state vector, $n$ the model order, $\mathcal{A}_c$ is the system matrix, $\mathcal{B}_c$ input matrix, $\mathcal{K}_c$ the input disturbance matrix (used later to implement Kalman filtering for state reconstruction), $C$ output matrix, $D$ feedforward matrix, $\mathcal{G}$ is the measurement noise matrix, $g_k$ is i.i.d. standard normal noise, $u$ input vector, ${\omega}$ a $n$-dimension standard Wiener process, and ${\theta}$ is the set of model parameters to estimate. Model parameters are estimated by using the \emph{greyest} function in MATLAB and minimizing the sum of the model one-step-ahead prediction errors. It was found that the model with best performance (namely, with uncorrelated model residuals) for the considered experimental BESS is a three time constant model ($n=3$) with structure as shown in Fig.~\ref{fig:modelstructure}.

The model is given by applying Kirchhoff laws to the circuit in Fig.~\ref{fig:modelstructure} and is:
\begin{align}
& {x}=\begin{bmatrix} v_{C_1} \;\; v_{C_2} \;\; v_{C_3}\end{bmatrix}, u_{tk} = \begin{bmatrix} i_{tk} \;\; 1 \end{bmatrix}^T \\
 & \mathcal{A}_c =
 \begin{bmatrix}
  \frac{-1}{R_1C_1} & \scriptstyle 0 & \scriptstyle 0 \\
  \scriptstyle 0 & \frac{-1}{R_2C_2} & \scriptstyle 0 \\
  \scriptstyle 0 & \scriptstyle 0 & \frac{-1}{R_3C_3}
 \end{bmatrix} ,
 \mathcal{B}_c = 
 \begin{bmatrix}
  \frac{1}{C_1} & \scriptstyle 0 \\
  \frac{1}{C_2} & \scriptstyle 0 \\
  \frac{1}{C_3} & \scriptstyle 0
 \end{bmatrix} \label{eq:systeminputmatrices}\\
 & \mathcal{K}_c=\text{diag}(k_1, k_2, k_3),\\
& \mathcal{C} =
 \begin{bmatrix}
1 \;\; 1 \;\; 1
\end{bmatrix},
\mathcal{D} =
\begin{bmatrix}
R_s \;\; E
\end{bmatrix}, \mathcal{G} = \sigma_g. \label{eq:voltage:statespaceCDG}
\end{align}
where  $R_1,C_1,R_2,C_2,R_3,C_3,k_1, k_2, k_3, R_s, E, \sigma_g$ is the set of parameters to estimate.
The estimated values of the system parameters according to the BESS SOC ranges are summarized in Table~\ref{tab:modelparameters}.

\begin{table}[!ht]
\renewcommand{\arraystretch}{1.4}
\centering
\caption{Estimated BESS voltage model parameters for different SOC ranges}\label{tab:modelparameters}
\begin{tabular}{| c|c|c|c|c|c|}
\hline
\bf SOC & \bf 0-20\% & \bf 20-40\% & \bf 40-50\% &  \bf 60-80\% &  \bf 80-100\% \\
\hline
$E$ & 592.2 & 625.0 & 652.9 & 680.2 & 733.2 \\
$R_s$ & 0.029 & 0.021 & 0.015 & 0.014 & 0.013 \\
$R_1$ & 0.095 & 0.075 & 0.090 & 0.079 & 0.199 \\
$C_1$ & 8930 & 9809 & 13996 & 9499 & 11234 \\
$R_2$ & 0.04 & 0.009 & 0.009 & 0.009 & 0.010 \\
$C_2$ & 909 & 2139 & 2482 & 2190 & 2505 \\
$R_3$ & 2.5e-3 & 4.9e-5 & 2.4e-4 & 6.8e-4 & 6.0e-4 \\
$C_3$ & 544.2 & 789.0 & 2959.7 & 100.2 & 6177.3 \\
$k_1$ & 0.639 & 0.677 & 0.617 & 0.547 & 0.795 \\
$k_2$ & -5.31 & -0.22 & -0.36 & -0.28 & 0.077 \\
$k_3$ & 5.41 & 40 & 0.40 & 2.83 & -0.24 \\
$\sigma^2$ & -1.31 & -0.42 & 0.3426 & 3.5784 & 2.7694 \\
\hline
\end{tabular}
\end{table}

\citefig{fig:modelA:residuals} shows the autocorrelation function (ACF) of the time series of model prediction errors (or model residuals), that is the difference between the training set of current/voltage measurements and the respective model predictions. Checking for model residuals correlation is a test widely adopted in system identification (see for example \cite{CTSM},\cite{}) to validate the ability of identified models of capturing the dynamics contained in the training data set. The test consists in comparing the ACFs of model residuals and white noise (which is i.i.d., therefore uncorrelated by definition): if all the components of the former ACF falls in the 95\% confidence interval of the latter, the test is successful and the dynamic performance of the model are considered satisfactory. The situation in \citefig{fig:modelA:residuals} refers to 50\% SOC and shows uncorrelated model residuals, thus denoting that the identified model is able to capture all dynamics contained in the training data set. Although not shown here for a reason of space, the same behavior was observed also for the other SOC ranges of Table~\ref{tab:modelparameters}.

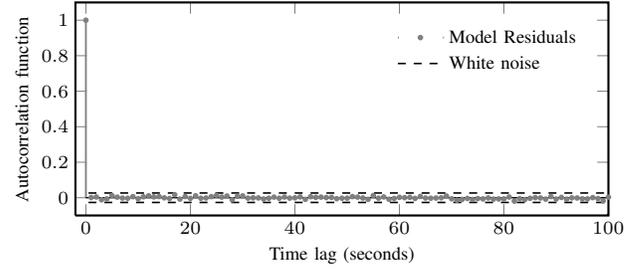
\begin{figure}[!ht]
\centering
\scriptsize
%
%

\pgfplotsset{major grid style={dotted,green!50!black}}

\begin{tikzpicture}

\begin{axis}[%
width=\matlabfigurewidth,
height=0.40\matlabfigurewidth,
at={(0\matlabfigurewidth,0\matlabfigurewidth)},
scale only axis,
xmin=-2,
xmax=100,
xlabel={Time lag (seconds)},
ymin=-0.1,
ymax=1.1,
ylabel={Autocorrelation function},
ytick={-0.2, 0, 0.2,0.4,0.6, 0.8, 1.0},
axis background/.style={fill=white},
legend style={at={(0.95,0.65)},anchor=south east,legend cell align=left,align=left,draw=none, fill=white, inner sep=0pt},
axis line style = {thick},
]
\addplot[ycomb,color=gray,solid,line width=0.7pt,mark size=0.7pt,mark=*,mark options={solid,fill=gray}] plot table[row sep=crcr] {%
0	1\\
1	0.00120723351396057\\
2	0.00411404110237128\\
3	-0.0114548486824811\\
4	-0.0065411887848084\\
5	0.0122413953080384\\
6	0.0033427272146689\\
7	-0.00190683640567353\\
8	-0.00295383249284927\\
9	0.00603457781604921\\
10	-0.00565602309630945\\
11	0.00523016749629586\\
12	0.0110549783800514\\
13	0.00472790124628741\\
14	0.00671724712933407\\
15	-0.000598579765112874\\
16	-0.004316533437702\\
17	0.0172852820221298\\
18	-0.00791236569469245\\
19	0.00692024799433014\\
20	-0.00345511810355915\\
21	0.0109142626487618\\
22	-0.00292414700081686\\
23	-0.00141087637471306\\
24	0.00636864119603651\\
25	0.0130414469915562\\
26	0.00355315506859097\\
27	0.00938221972218566\\
28	-0.0115908381027235\\
29	0.00831394837044916\\
30	0.0103935952749029\\
31	-0.00209588955804041\\
32	-0.000579230465375833\\
33	-0.00119901947239531\\
34	-0.00726523578446813\\
35	-0.00282788746511753\\
36	0.00341355286442482\\
37	-0.00291570517467642\\
38	0.00388973718520851\\
39	-0.00236982703633526\\
40	-0.000468109590337211\\
41	0.00463038363009724\\
42	-0.00913849882176222\\
43	0.00569817787893118\\
44	0.00216313966515816\\
45	0.00340812701340863\\
46	0.000994653758323909\\
47	0.00200552744352019\\
48	-0.00542190920139603\\
49	-0.00270594790663863\\
50	-0.00248504640259393\\
51	0.00644959687164941\\
52	0.00355523467096577\\
53	-0.00142873257133979\\
54	-0.00933975192618052\\
55	0.0121975080084497\\
56	-0.00235111026035316\\
57	0.00446493016653827\\
58	-0.00868405589715571\\
59	-0.00412055213371298\\
60	0.0023349786090473\\
61	0.000276683884771881\\
62	0.00235215869862704\\
63	-0.00409772866112891\\
64	0.00712145725719424\\
65	-0.00164213646259702\\
66	-0.00286695209646637\\
67	-0.00184348989671778\\
68	-0.000219693836106584\\
69	0.00974876775252358\\
70	-0.00606735307750222\\
71	-0.014140961434949\\
72	-0.00960156826548973\\
73	-0.00420406306623778\\
74	-0.010026578791516\\
75	-0.00481643309063189\\
76	-0.00666037681912679\\
77	-0.00108847096796747\\
78	-0.00787659009352406\\
79	-0.00674520798296407\\
80	-0.00669345296550911\\
81	0.00377685227313802\\
82	-0.0189360353833839\\
83	-0.00528000265570119\\
84	-0.0105356800797426\\
85	-0.00339523795146258\\
86	0.000724909061977572\\
87	-0.00130943726515911\\
88	-0.00792747270851331\\
89	0.00679090661037315\\
90	-0.00733604476395395\\
91	0.00166868977579069\\
92	-0.00819033480597336\\
93	-0.000745885522165842\\
94	-0.00159148352872096\\
95	-0.00619497561755806\\
96	-0.00646505898715999\\
97	0.000536933527923076\\
98	-0.00467438838700466\\
99	-0.0178879217586231\\
100	0.00313248494732545\\
};
\addlegendentry{Model Residuals};

\addplot [color=black,dashed,line width=0.6pt,forget plot]
  table[row sep=crcr]{%
0.5	0.0266477239069211\\
100	0.0266477239069211\\
};
\addplot [color=black,dashed,line width=0.6pt]
  table[row sep=crcr]{%
0.5	-0.0266477239069211\\
100	-0.0266477239069211\\
};
\addlegendentry{White noise};

\addplot [color=black,solid,line width=0.5pt,forget plot]
  table[row sep=crcr]{%
0	0\\
100	0\\
};
\end{axis}
\end{tikzpicture}%
\caption{Autocorrelation function (ACF) of model residuals (full line) and white noise (horizontal lines) at 95\% confidence level.}
\label{fig:modelA:residuals}
\end{figure}

\paragraph{Model integration in MPC}
The continuous time model in  \eqref{eq:ssd:1} is discretized at $T_s=10$~seconds resolution in order to be implemented in the MPC constraints in \eqref{eq:rt:mpc3}. Discretization is performed with the forward difference (or forward Euler). However, in order not to incur in numerical instability, the quickest time constant given by the $(R_3,C_3)$ branch is dropped in favor of an algebraic state by using the matched DC gain method. Let $\mathcal{A}_r, \mathcal{B}_r, \mathcal{K}_r$ be the continuous time system, input and system noise matrices of the reduced order model, the discretized system and input matrices are given by:
\begin{align}
 & \mathcal{A} = {1} + \mathcal{A}_r T_s \label{eq:dmodel00}\\
 & \mathcal{B} = \mathcal{B}_r T_s \\
 & \mathcal{K} = \mathcal{M}_{1} \mathcal{M}_{2}^T \label{eq:dmodel01},
\end{align}
where $\mathcal{M}_{1}$ and $\mathcal{M}_{2}$ are given according to the Van Loan's method \cite{1101743} and are matrix exponentials, which are approximated by the first order truncation of their respective Taylor expansions:
\begin{align}
&\mathcal{M}_1 = e^{\mathcal{K}_rT_s} = {1} + \mathcal{K}_rT_s\\
&\mathcal{M}_2 = e^{A_r^TT_s} = {1} + \mathcal{A}_r^TT_s.
\end{align}
The matrices $\mathcal{A}, \mathcal{B}, \mathcal{K}$ in \eqref{eq:dmodel00}-\eqref{eq:dmodel01} and $\mathcal{C}, \mathcal{D}, \mathcal{G}$ in \eqref{eq:voltage:statespaceCDG} constitute the skeleton of the discrete time state-space  of the BESS equivalent circuit model. By using the procedure shown in Appendix \ref{sec:MPCtransitionmatrices}, they are used to generate the transition matrices $\phi^v, \psi_i^v, \psi^v_1$, which are finally used to implement the MPC voltage constraints in \eqref{eq:rt:mpc3}.

\paragraph{State estimation}
The components of the system state in \eqref{eq:ssd:1} are a modelling abstraction and cannot be measured. However, their value must be known in order to compute the BESS voltage predictions. In turns, the state is estimated from measurements of the battery DC voltage by applying Kalman filtering (KF, \cite{STENGEL}). As known, it consists in a two-stage procedure, repeated at each discrete time interval: a prediction step to determine the system evolution (state expected value and covariance matrix $P$) solely on the basis of the knowledge on the system
\begin{align}
 {{x}}_{k|k-1} &= \mathcal{A} {{x}}_{k-1|k-1} + \mathcal{B} {u}_{k-1} \label{eq:kalman:prediction}\\
 P_{k|k-1} &= \mathcal{A} P_{k-1|k-1} \mathcal{A}^T + \mathcal{K}\mathcal{K}^T \label{eq:kf:sc},
\end{align}
and an update stage, where the predicted state is corrected accounting for the last measurement $v_k$
\begin{align}
 {{x}}_{k|k} &= {{x}}_{k|k-1} + G (y_k - \mathcal{C} {{x}}_{k|k-1}) \label{eq:kf:u1} \\
 P_{k|k} &= \left(P_{k|k-1}^{-1} + \mathcal{C}^T\sigma_g^{-1}\mathcal{C}\right)^{-1}. \label{eq:kf:u2}
\end{align}
where $G$ is the Kalman gain:
\begin{align}
 G = P_{k|k-1}\mathcal{C}^T\left( \mathcal{C} P_{k|k-1} \mathcal{C}^T + \sigma_g^2\right)^{-1} \label{eq:kf:kg},
\end{align}
and $\sigma_g$ is the measurement noise (known from the parameters estimation).
KF requires full system observability, that in our case is enforced by construction since the model is estimated from measurements. 

\subsubsection{BESS SOC} \label{eq:pmodel:soc}
The BESS SOC model is:
\begin{align}
 \text{SOC}_{k+1} = \text{SOC}_{k} + \frac{10}{3600}\frac{i_k}{C_\text{nom}} \label{eq:socmodel},
\end{align}
where $C_\text{nom}=\SI{810}{\ampere\hour}$ (ampere-hour) is BESS capacity from datasheet information. It is worth noting that we neglect here the dependency of the BESS capacity with respect to the C-rate and cells temperature, see for example \cite{belvedere2012microcontroller}. The discretized state-space matrix are easily found from \eqref{eq:socmodel} and are $A=1,\ B=10/3600/C_\text{nom},\ C=1,\ D=0$. They are used as shown in Appendix \ref{sec:MPCtransitionmatrices} to generate the transition matrices $\phi^\text{SOC}, \psi_i^\text{SOC}$ for the SOC predictive constraints in \eqref{eq:rt:mpc4}.

\subsection{On the use of BESS models in the day-ahead and real-time stages}
Open-loop constraints on BESS operation are enforced in both day-ahead and real-time stages. The main differences between the two implementations are the decision variables, the time resolution at which they are computed (5~minutes and 10~s, respectively), and the fact that the latter problem is re-evaluated at each time interval by incorporating new information from real-time measurements.
From the point of view of their practical purpose, BESS models are integrated with the objective of modeling the storage capacity and constraints. However, while in the day-ahead formulation they are to represent the long-term flexibility, in real-time operation they model short-term operational constraints, therefore enabling the computation of reliable control actions which are respectful of the BESS operational limits. This explains the reason why more accurate BESS models are implemented in the latter stage than in the former, where a high level of details is not needed, especially considering that a more conservative plan for the BESS usage can be achieved by adding back-off terms to SOC and power flow constraints in \eqref{eq:dayahead:ic0}-\eqref{eq:dayahead:ic3}.

\subsection{Implementation of the real-time strategy}\label{sec:realtimeimplementation}
\begin{figure}[ht]
\centering {
\scriptsize
\tikzstyle{decision} = [diamond, draw, fill=blue!0, 
    text width=4.5em, text badly centered, node distance=3cm, inner sep=0pt]
\tikzstyle{block} = [rectangle, draw, fill=gray!10, 
    text width=15em, text centered, rounded corners, minimum height=3em]
\tikzstyle{line} = [draw, -latex']
    
\begin{tikzpicture}[node distance = 1.25cm, auto]
    \node [block, text width=28em] (init) {00:00:00 UTC (k=0, beginning of the day of operation)\\ Retrieve dispatch plan $\widehat{P}$ from database};
    \node [block, below of=init, text width=28em] (cc) {Realizations of the power flow at the GCP $P_{k-1}$ and the BESS demand $B_{k-1}$ for the previous 10~s time interval become known from measurements.};
    
    \node [block, below of=cc, text width=28em] (pmu) {Retrieve dispatch plan set-point $P^*_k$, $\underline{k}$ and $\overline{k}$ by applying \eqref{eq:rt:setpoint}-\eqref{eq:rt:uplimit}};
    
    \node [block, below of=pmu, text width=25em] (computeerror) {Compute the short-term prosumption predictions and \\used them together with the available information from $\underline{k}$ to $k$ to determine the dispatch plan error $e_k$ in \eqref{eq:rt:dpe}.};
    
    \node [block, below of=computeerror,text width=28em] (evaluate) {Read $\text{SOC}_k$ and $v_k$ from the BMS, and update Kalman filtering};

    \node [block, below of=evaluate, text width=28em] (solve) {Select the right BESS voltage model according to $\text{SOC}_k$ (model scheduling) ,\\compute the short-term prosumption forecast, and \\ determine $\boldsymbol{i}^o_{k|\overline{k}}$ by solving \eqref{eq:rt:mpc0}-\eqref{eq:rt:mpc6} for the time horizon $k$ to $\overline{k}$};
    
    \node [block, below of=solve, text width=28em] (apply) {Extract the first current value from the control law, find the BESS real power set-point with \eqref{eq:powersetpoint} and send it to the BMS for actuation.};
    
    \node [block, below of=apply, text width=28em] (wait) {Wait for 10~seconds};
    
    \node [block, left of=computeerror, node distance=4cm, text width=5em] (next) {$k=k+1$};
    
    \node [block, below of=wait, text width=28em] (stop) {Stop at 24:00:00~UTC};
    
    \path [line] (init) -- (cc);
    \path [line] (cc) -- (pmu);
    \path [line] (pmu) -- (computeerror);
    \path [line] (computeerror) -- (evaluate);
    \path [line] (evaluate) -- (solve);
    \path [line] (solve) -- (apply);
    \path [line] (apply) -- (wait);
    \path [line] (wait) -| (next);
    \path [line] (next) |- (cc);
    \path [line] (wait) -- (stop);
\end{tikzpicture}
}
\caption{Flow chart showing real-time operation during 24~hours.}\label{fig:flowchart}
\end{figure}
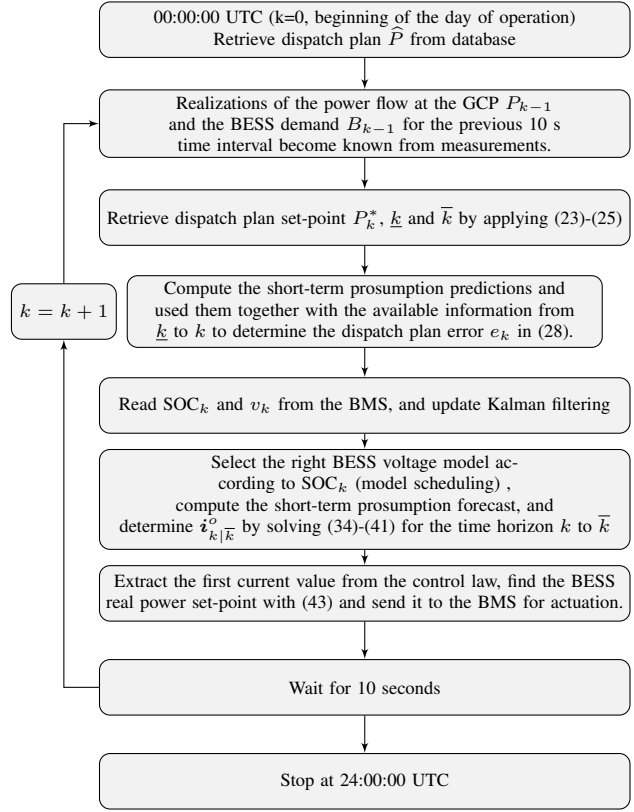

The flow of operation during the real-time strategy is sketched in Figure~\ref{fig:flowchart}. Real-time operation is implemented as a Matlab script and executed daily on the same computer as the day-ahead strategy. Also, information which become available from real-time measurements (power flow at the GCP, DC voltage, DC current, SOC, AC power flow of the BESS) are acquired and stored in a time series database with 1~second resolution.

\begin{figure}[!ht]
\begin{center} 
\includegraphics[width=0.9\columnwidth]{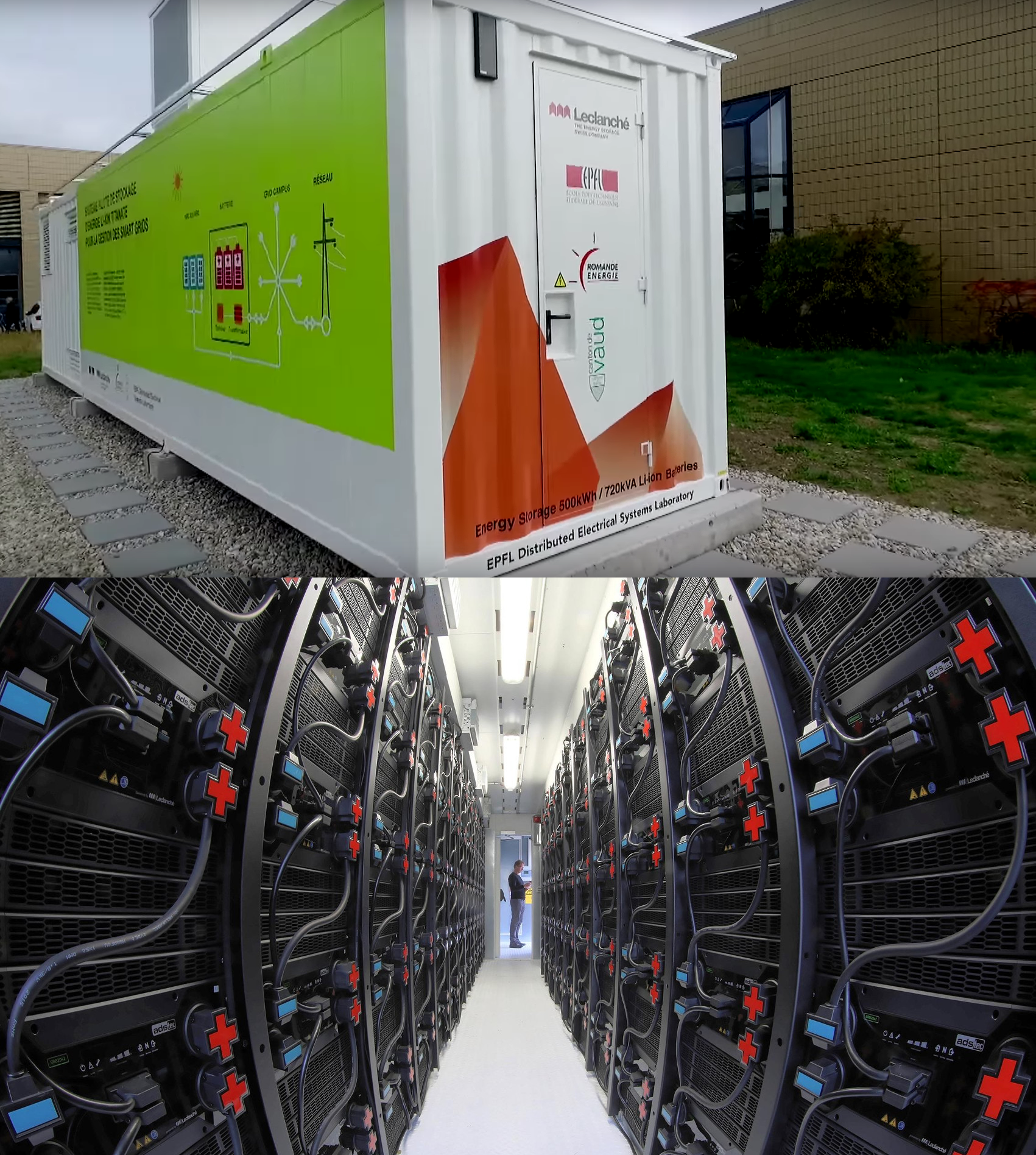}
\end{center}
\vspace{-3mm}
\caption{An outside (top picture) and inside view (bottom picture) of the 720~kVA/500~kWh Lithium Titanate-based BESS used in this experimental work. The system, developed by Leclanché, includes a four-quadrant fully controllable DC/AC converter and a 0.3/20~kV step-up transformer. It is hosted in a temperature controlled container (picture of Alain Herzog, EPFL).}\label{fig:container}
\end{figure}

As introduced in Section~\ref{sec:statement}, experiments are performed on a real-scale real-life MV distribution grid using a grid-connected 720~kVA/500~kWh BESS. The BESS is based on the Lithium titanate technology and is rated for 15 thousands cycles. It consists in 9 parallel racks (each is composed by 15 modules in series, and each module is composed by a 20s3p cell pack), a four quadrant power converter, everything installed in a temperature controlled container. A view of the experimental BESS is shown in Fig.~\ref{fig:container}.

\section{Experimental Results and Discussion}\label{sec:results}
\begin{figure} [!t]
\centering

\subfloat[\scriptsize \bf Day-ahead: prosumption uncertainty sets and expected value, and offset plan.] {
\scriptsize \input{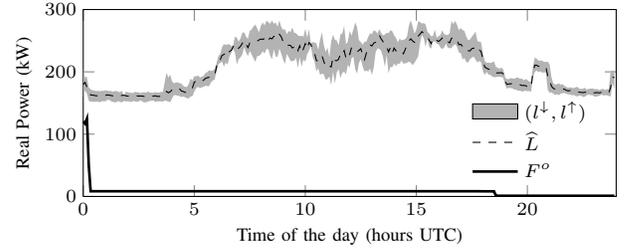}
\label{fig:day0_a}}

\vspace{-2mm}
\subfloat[\scriptsize \bf GHI forecast \emph{vs} realization and respective average components.] {
\scriptsize \input{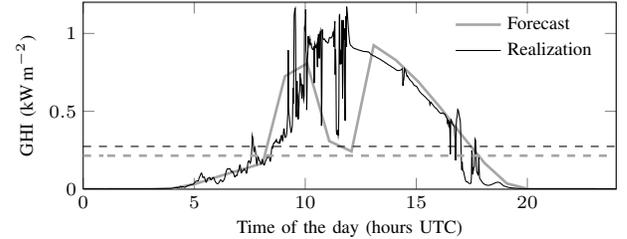}
\label{fig:day0_b}}

\vspace{-2mm}
\subfloat[\scriptsize \bf Real-time: \emph{dispatch plan} \emph{vs} realization of GCP power transit and prosumption.] {
\scriptsize \input{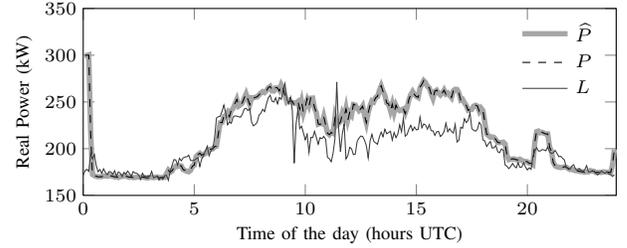}
\label{fig:day0_c}}

\vspace{-2mm}
\subfloat[\scriptsize \bf Realizations of BESS SOC, DC current and voltage and respective limits.] {
\scriptsize \input{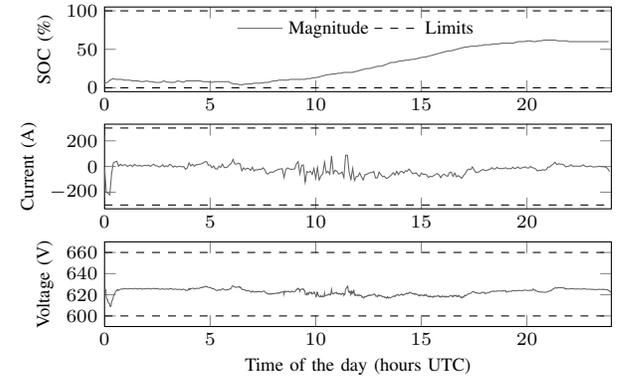}
\label{fig:day0_d}}

\caption{Day 0: experimental results.} \label{fig:day0}

\end{figure}

\begin{figure} [!t]
\centering

\subfloat[\scriptsize \bf Day-ahead: prosumption uncertainty sets and expected value, and offset plan.] {
\scriptsize \input{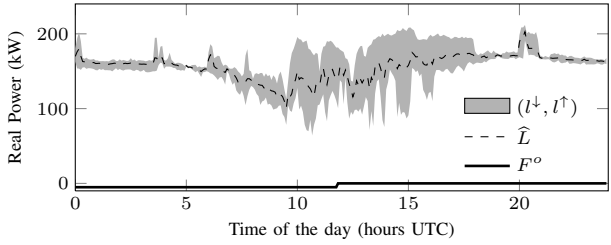}
\label{fig:day1_a}}

\vspace{-2mm}
\subfloat[\scriptsize \bf GHI forecast \emph{vs} realization and respective average components.] {
\scriptsize \input{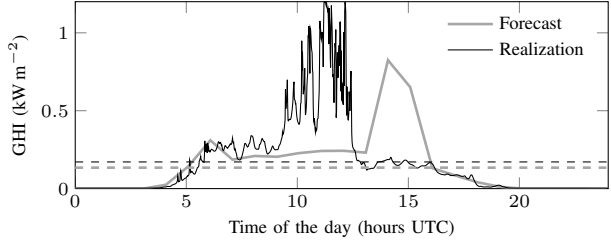}
\label{fig:day1_b}}

\vspace{-2mm}
\subfloat[\scriptsize \bf Real-time: \emph{dispatch plan} \emph{vs} realization of GCP power transit and prosumption.] {
\scriptsize \input{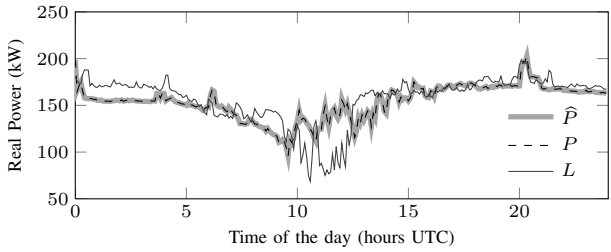}
\label{fig:day1_c}}

\vspace{-2mm}
\subfloat[\scriptsize \bf Realizations of BESS SOC, DC current and voltage and respective limits.] {
\scriptsize \input{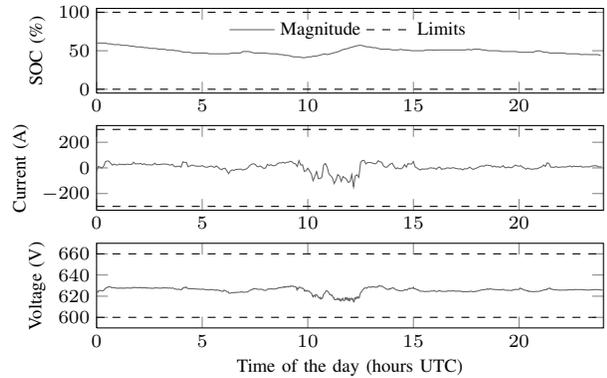}
\label{fig:day1_d}}

\caption{Day 1: experimental results.} \label{fig:day1}
\end{figure}

\begin{figure} [t]
\centering

\subfloat[\scriptsize \bf Day-ahead: prosumption uncertainty sets and expected value, and offset plan.] {
\scriptsize \input{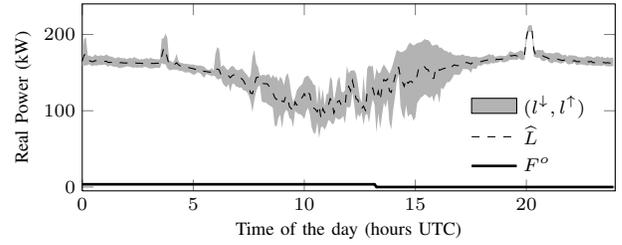}
\label{fig:day2_a}}

\vspace{-2mm}
\subfloat[\scriptsize \bf GHI forecast \emph{vs} realization and respective average components.] {
\scriptsize \input{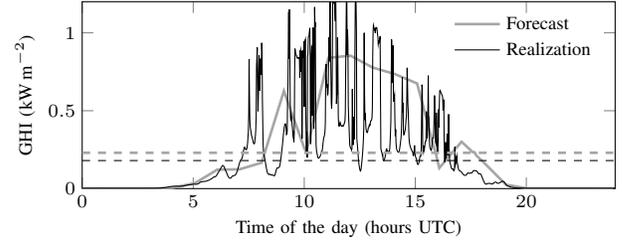}
\label{fig:day2_b}}

\vspace{-2mm}
\subfloat[\scriptsize \bf Real-time: \emph{dispatch plan} \emph{vs} realization of GCP power transit and prosumption.] {
\scriptsize \input{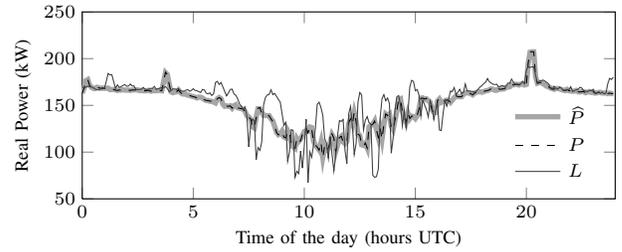}
\label{fig:day2_c}}

\vspace{-2mm}
\subfloat[\scriptsize \bf Realizations of BESS SOC, DC current and voltage and respective limits.] {
\scriptsize \input{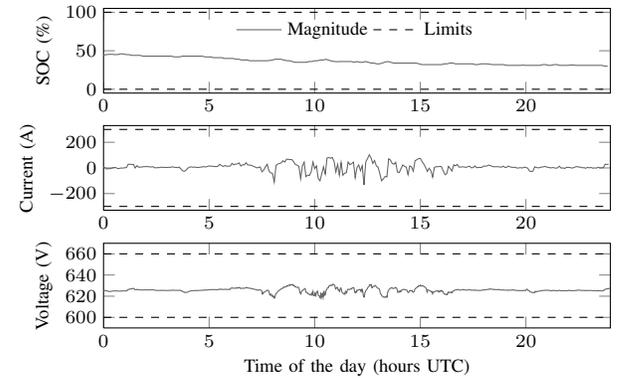}
\label{fig:day2_d}}

\caption{Day 2: experimental results.} \label{fig:day2}
\end{figure}

In this section, we show the experimental results obtained by dispatching the operation of the MV feeder described in Section~\ref{sec:statement}.
We start by describing three contiguous days of operation (denoted by day~0, day~1 and day~2) in order to illustrate the ability of the proposed method to manage the SOC and achieve continuous time operation. Results are shown in figures \ref{fig:day0}-\ref{fig:day2}. Each figure consists in four main plots \emph{a, b, c, d}, which respectively shows:

\begin{itemize}
 \item[(a)] the extreme values of the prosumption uncertainty sets $\left(l^\downarrow_i, l^\uparrow_i\right)$, the expected prosumption $\widehat{L}_i, i=0,\dots, N-1$, and the offset profile $F^o$. The first two quantities are the outcome of the forecasting tool discussed in \ref{sec:ppredictions}, while the last is computed as detailed \ref{sec:offsetplan}. The second and third quantities are used to generate the \emph{dispatch plan} according to \eqref{eq:dispatchplan} $\widehat{P}_i$;
 
 \item[(b)] the GHI day-ahead forecast, the respective realizations, respective average components (denoted with the dashed lines). As described in \ref{sec:ppredictions}, the latter quantity is used to determine the total radiation in one day and select the most suitable prosumption profiles among the historical data;
 
 \item[(c)] the \emph{dispatch plan} $\widehat{P}_i$, the power transit at the GCP realizations $P_i$ and the prosumption realizations $L_i$;
 
 \item[(d)] the measurements of the BESS SOC, DC current $i$ and voltage $v$ and their respective limits.
 
\end{itemize}

As visible from Fig.~\ref{fig:day0_d}, the BESS starts the operation in the day~0 experiment with a very low SOC. In general (and assuming for one moment unbiased prosumption forecast), this is not a desirable situation because the BESS capacity of absorbing forecasting errors is not symmetric. The specific objective of the offset profile is to restore a sufficient BESS SOC in order that enough BESS energy capacity is available along the day to compensate for forecasting errors, which are modeled in the optimization problem in \eqref{eq:dayahead:cost}-\eqref{eq:dayahead:ic4} by the prosumption uncertainty sets (shown in Fig.~\ref{fig:day0_a} with the shaded band). In this case, the offset profile is such that the \emph{dispatch plan} overestimates the prosumption, therefore implicitly achieving the BESS to slowly charge thanks to absorbing the increased level of demand. Figure \ref{fig:day0_b} shows that day ahead GHI forecast are fairly accurate in terms of average components, which is what is considered in the adopted forecasting algorithm. As visible from Fig.~\ref{fig:day0_c}, during operation the power flow at the GCP $P$ follows precisely the \emph{dispatch plan} $\widehat{P}$, thus denoting the good tracking performance of the MPC. A more accurate numerical comparison is shown later in this section. Fig.~\ref{fig:day0_d} shows the details of BESS operation. As described in the paper, the BESS is controlled in order to compensate for the mismatch between prosumption realization and dispatch plan, and, in this case, it slowly charges along the day, as imposed by the offset plan. The BESS constraints, which are enforced in the MPC are generally respected during operation.

The BESS terminates the day~0 experiment with a SOC above the 50\% level. This is reflected on the operation of the next day (experiment day~1, in Fig.~\ref{fig:day1}), which is characterized by a slightly negative offset plan ($F^o$ in Fig.~\ref{fig:day1_a}) to restore a lower a BESS SOC, as visible in Fig.~\ref{fig:day1_d}. As shown in Fig.~\ref{fig:day1_c}, the algorithm achieves to track the \emph{dispatch plan} successfully. As visible by comparing figures \ref{fig:day0_d} and \ref{fig:day1_d}, the BESS SOC variation in the latter case is smaller than in the former, thus denoting a smaller accumulated forecasting error.

Fig.~\ref{fig:day2} shows the operation of the experiment on day~3. In this case, the BESS SOC is below the 50\% level since the previous day of operation, and a positive offset profile $F^o$ (Fig.~\ref{fig:day2_a}) is necessary to raise the SOC. During operation, the algorithm is able to control the BESS in order to track the dispatch plan successfully (Fig.~\ref{fig:day2_c}). However, the BESS SOC slowly drifts away from the 50\% level during the day, as visible in Fig.~\ref{fig:day2_d}, to compensate for a moderate overestimation of the average PV production (as visible in Fig.~\ref{fig:day2_b}).

Table~\ref{tab:performance} summarizes the tracking performance of the proposed control strategy in the experimental results. For each day experiment, two cases are considered:
\begin{itemize}
 \item {\bf no dispatch}: this corresponds to nowadays conventional operation, namely when prosumption is not dispatched but simply forecasted (although on a higher aggregation level than as done here). In this context, the tracking error is defined as the difference between the prosumption forecast and the prosumption realization;
 \item {\bf dispatch}: the operation of the feeder is dispatched as described previously. The error here is given by the difference between the dispatch plan and the power flow realizations at the GCP.
\end{itemize}
Three metrics are considered: the root mean square error (RMSE) along with its mean and maximum absolute value. As visible from Table~\ref{tab:performance}, dispatched operation achieve better figures than the base case. The tracking performance of the MPC are fairly accurate, with an RMS value less than \SI{0.5}{\kilo\watt}, which is approximately the 0.2\% of the feeder average power consumption (200~kW).

\begin{table}[!ht]
\vspace{-0mm}
\renewcommand{\arraystretch}{1.2}
\centering
\caption{Tracking Error Statistics (in kW) for Experiments Day~0 to 2}\label{tab:performance}
\begin{tabular}{| l | C{1.5cm} |C{1.0cm}|C{1.0cm}|C{1.0cm}|}
\hline
\multicolumn{2}{|c|}{Experiment} & RMSE & Mean & Max \\
\hline
\multirow{ 2}{*}{Day~0}  & no dispatch & 19.20 & 4.68 & 50.82\\
& dispatch & 0.43 & $< 0.01$ & 1.54 \\
\hline
\multirow{2}{*}{Day~1} & no dispatch & 18.71 & -0.39  & 72.150\\
 & dispatch & 0.25 & $< 0.01$ & 0.740 \\
\hline
\multirow{2}{*}{Day~2} & no dispatch & 18.06 & -4.92  & 54.45\\
 & dispatch & 0.42 & $< 0.01$ & 1.41 \\
\hline
\end{tabular}
\end{table}

\begin{figure} [!t]
\centering

\subfloat[\scriptsize \bf Day-ahead: prosumption uncertainty sets and expected value, and offset plan.] {
\scriptsize \input{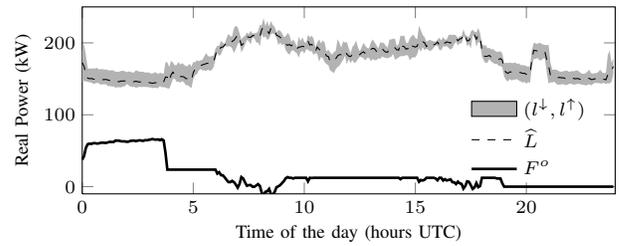}
\label{fig:day3_a}}

\vspace{-3mm}
\subfloat[\scriptsize \bf GHI forecast \emph{vs} realization and respective average components.] {
\scriptsize \input{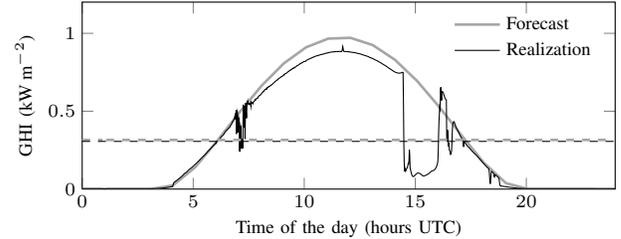}
\label{fig:day3_b}}

\vspace{-3mm}
\subfloat[\scriptsize \bf Real-time: \emph{dispatch plan} \emph{vs} realization of GCP power transit and prosumption.] {
\scriptsize \input{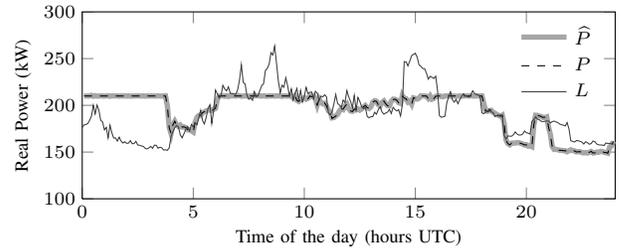}
\label{fig:day3_c}}

\vspace{-3mm}
\subfloat[\scriptsize \bf Realizations of BESS SOC, DC current and voltage and respective limits.] {
\scriptsize \input{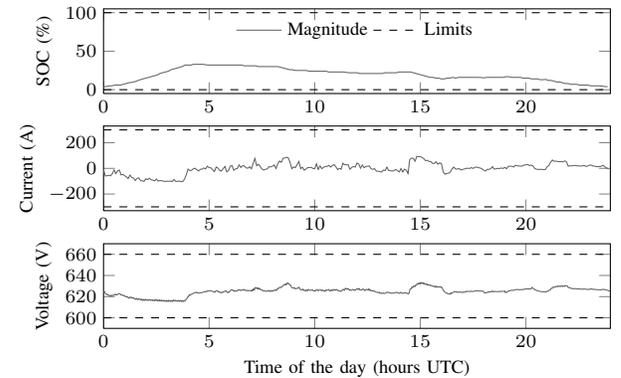}
\label{fig:day3_d}}

\caption{Day 3: experimental results.} \label{fig:day3}
\end{figure}

Finally, Fig.~\ref{fig:day3} shows a fourth day of operation where the parameter $P_\text{max}$ in \eqref{eq:dayahead:ic4} has been set to 210~kW to explicitly shows the capability of performing peak shaving. As visible from Fig.~\ref{fig:day3_c}, the magnitude of the  \emph{dispatch plan} is limited to the aforementioned value. Although the prosumption realization has peaks of 250~kW, the power flow at the GCP sticks to the dispatch plan, therefore achieving consumption peak shaving as well as dispatched operation. 

Table~\ref{tab:performance:a} summarizes in numerical form the operation during the experiment on day~3.

\begin{table}[!ht]
\vspace{-0mm}
\renewcommand{\arraystretch}{1.2}
\centering
\caption{Tracking Error Statistics (in kW) for Experiment on Day~3}\label{tab:performance:a}
\begin{tabular}{| l | C{1.5cm} |C{1.0cm}|C{1.0cm}|C{1.0cm}|}
\hline
\multicolumn{2}{|c|}{Experiment} & RMS & Mean & Max \\
\hline
\multirow{ 2}{*}{Day~3}  & no dispatch & 20.579 & -5.451 & 59.270\\
& dispatch & 0.196 & $< 0.01$ & 0.570 \\
\hline
\end{tabular}
\end{table}

Finally, Table~\ref{tab:time} shows statistics on the computation time required to complete the day-ahead and real-time procedures. The day-ahead procedure consists in performing the day-ahead prosumption forecasting (which is the most time consuming operation in this phase because it requires to load a large amount of historical data), computing the offset profile by solving the convex optimization problem \eqref{eq:dayahead:cost}-\eqref{eq:dayahead:ic4}, and determining the \emph{dispatch plan} with \eqref{eq:dispatchplan}. The real-time procedure consists in updating the Kalman filtering for reconstructing the state of the BESS voltage model, computing the short-term forecast of the prosumption  and solving the convex optimization problem in \eqref{eq:rt:mpc0}-\eqref{eq:rt:mpc6} to determine the BESS current trajectory. The computation is performed in Matlab on a Debian Linux machine equipped with an Intel i5 processor and 4Gb of RAM. The computation time of both procedures is quite short. In particular, real-time computation (which is the most time critical among the two because it needs to be recomputed every 10~s to determine the control decision) is solved in less than half a second, considerably less than the 10~s period at which is actuated.

The statistics in Table~\ref{tab:time} do not include the communication time, which however is as the typical low latency of local are network (in the range of tens of milliseconds, see for example \cite{EPFL-CONF-203775}).

\begin{table}[!ht]
\vspace{-0mm}
\renewcommand{\arraystretch}{1.2}
\centering
\caption{Computation times}\label{tab:time}
\begin{tabular}{| l | C{1.5cm} | C{1.75cm} | }
\hline
Procedure & Mean (second) & Standard Deviation (second) \\
\hline
Day-ahead problem & 6.5 & 0.2  \\
Real-time problem & 0.55 & $< 0.1$ \\
\hline
\end{tabular}
\end{table}

\section{Conclusions and perspectives}\label{sec:concs}
We have discussed a control framework to dispatch the operation of a cluster of stochastic prosumers according to a profile that is established the day before the operation by using a BESS as a controllable element. It consists of a two-stage procedure: day-ahead scheduling and real-time operation. In the former phase, the \emph{dispatch plan} (namely the prosumption profile at 5 minutes resolution that the feeder should follow during operation) is determined as the sum between the prosumption forecasted profile and the so-called offset profile, which allows restoring an adequate level of BESS flexibility for the next day of operation. In the latter phase, MPC is used to compensate for the mismatch between prosumption realization and dispatch plan.

Prediction models that are applied in the problem (prosumption forecasting models and BESS voltage models) are identified from measurements applying data-driven approaches. The optimization problems which were developed as a part of the control framework are all formulated as convex ones. Therefore, they are characterized by an enhanced level of tractability and efficient to solve. In particular, the BESS MPC allows formulating the BESS energy throughput in the objective function while retaining the linearity of the expressions of the BESS SOC, DC voltage and current constraints.

The control framework is validated on a real-life real-scale MV system of the EPFL campus, the so-called \emph{dispatchable feeder}, which consists in a distribution feeder interfacing a number of office buildings (300~kW peak demand) equipped with roof-top PV generation (90~kWp). The controllable element is a grid-connected 720~kVA-500~kWh Lithium titanate battery. The experimental results show that the proposed control framework is able to dispatch the operation of the group of prosumers along multiple days of operation with a good tracking performance (the tracking error RMS value is 0.5~kW and the mean is below 0.01~kW with 200~kW average prosumption).

The proposed strategy can be regarded as a solution for the integration of utility-scale BESSs. It allows to achieve dispatched operation of specific MV nodes by design, while relying on a minimally invasive monitoring and control infrastructure (only the BESS and the power flow at the GCP are required) and with minimal coordination requirements with the upper grid layer because all the complexity is masked behind the commitment of the operator to track the dispatch plan. Future works concern the improvement of forecasting tools performance for low levels of aggregation, the definition of intra-day mechanisms to manage contingencies (for example, when the storage capacity is saturated), the establishment of planning criteria to size the storage and economic assessment of the proposed strategy compared to the conventional way of regulating power procurement.

\appendices

\section{Convex Formulation of Day-ahead Problem}\label{appendix:convexformulation}
The formulation in \eqref{eq:dayahead:cost}-\eqref{eq:dayahead:ic4} is nonconvex due to the relationship \eqref{eq:conversion:efficiency}, which involves a nonlinear function of the decision variable. We apply an augmented formulation of what proposed in \cite{kraning2011operation} to obtain a convex equivalent formulation of the original problem.  We introduce:
\begin{align}
 K_i = F^o_i + L^\downarrow_i \label{eq:appAR1:K}\\
 G_i = F^o_i + L^\uparrow_i \label{eq:appAR1:G}
\end{align}
which we decompose into their respective positive and strictly negative parts:
\begin{align}
 & K_i = K_i^+ - K_i^-, && K_i^+\ge0,\ K_i^-<0\\
 & G_i = G_i^+ - G_i^-, && G_i^+\ge0,\ G_i^-<0.
\end{align}
The quantities $K_i^+,K_i^-, G_i^+,G_i^-, i=0,\dots,N-1$ become the decision variables of the problem. They are coupled by the relationships \eqref{eq:appAR1:K}-\eqref{eq:appAR1:G} because the offset plan $F^o_i$ must be the same in the two cases. The coupling equality constraint is given by determining $F^o_i$ in \eqref{eq:appAR1:K}-\eqref{eq:appAR1:G} and imposing $F_i^o$ to be the same in \eqref{eq:appAR1:K} and \eqref{eq:appAR1:G}, which gives:
\begin{align}
K_i^+ - K_i^- - L^\downarrow_i = G_i^+ - G_i^- - L^\uparrow_i .
\end{align}

The quantity ${K_i^+}^o, {K_i^-}^o, {G_i^+}^o, {G_i^-}^o,\ i=0,\dots, N-1$ are finally given by the following linear (convex) program:
\begin{align}
 \underset{ 
 \renewcommand*{\arraystretch}{0.8}
 \begin{matrix}
  \scriptstyle \boldsymbol{K^+,K^-, G^+,G^-} \in \mathbb{R}^{N} \\
 \end{matrix}}
 {\text{arg~min}}~\Bigg\{
 \sum_{i=1}^{N} \big({K_i^+} + {K_i^-}  + {G_i^+} + {G_i^-} \big)
\Bigg\}\label{eq:dayahead:convex0}
\end{align}
subject to:
\begin{align}
 & K_i^+ \ge 0 \\
 & K_i^- \ge 0 \\
 & G_i^+ \ge 0 \\
 & G_i^- \ge 0 \\
 & K_i^+ - K_i^- - L^\downarrow_i = G_i^+ - G_i^- - L^\uparrow_i \\
 & \text{SOE}^\downarrow_{i+1} = \text{SOE}^\downarrow_{i} + \beta^+ K_i^+ - \beta^- K_i^- \label{eq:soedown:linear} \\
 & \text{SOE}^\uparrow_{i+1} = \text{SOE}^\uparrow_{i} + \beta^+ G_i^+ - \beta^- G_i^-. \label{eq:soeup:linear} \\
 & \text{SOE}^\uparrow_{i} \ge \text{SOE}_\text{min}  \label{eq:dayahead:convex8} \\
 & \text{SOE}^\uparrow_{i} \le \text{SOE}_\text{max}  \label{eq:dayahead:convex9} \\
 & {K_i^+} - {K_i^-} \le B_\text{max}  \label{eq:dayahead:convex10} \\
 & {K_i^+} - {K_i^-} \ge B_\text{min}  \label{eq:dayahead:convex11} \\
 & {G_i^+} - {G_i^-} \le B_\text{max}   \label{eq:dayahead:convex12} \\
 & {G_i^+} - {G_i^-} \ge B_\text{min}  \label{eq:dayahead:convex13}
\end{align}
for $i=0,1\dots,N-1$. The dispatch plan is finally:
\begin{align}
 F^o_i  = {K_i^+}^o - {K_i^-}^o - L^\downarrow_i, && i=0,\dots,N-1 \label{eq:determining_offset_plan}.
\end{align}
We also note that the relationship above is also linear, thus it can be used to formulate additional constraints for the offset plan (to implement for example peak shaving).

\section{Derivation of the Transition Matrices for MPC}\label{sec:MPCtransitionmatrices}
We consider a linear dynamic model with the following discrete state-space representation (as those developed in Section~\ref{sec:predictions} for the BESS SOC and voltage):
\begin{align}
 & x_{k+1} = Ax_k + Bu_k \label{eq:app:statespace:0}\\
 & y_k = C x_k + D u_k \label{eq:app:statespace:1}
\end{align}
where $x_k \in \mathbb{R}^n$ is the state vector at discrete time interval $k$, ${u}_k \in \mathbb{R}$ is the input, $y \in \mathbb{R}$ is the system output, $A$ is the $n\times n$ system matrix, $B$ is the $n\times 1$ input matrix, $C$ is the $1\times n$ output matrix, and the scalar $D$ is the feed-forward gain. For the moment, we consider the case with only one input signal: the extension to multiple inputs is shown at the end of this section. The evolution of the state vector ${x}$ from a known initial state ${x}_0$ as a function of a given input sequence $u_0, u_1, \dots, u_N$ using \eqref{eq:app:statespace:0} is
\begin{align}
& {x}_1 = A{x}_0 + B {u}_0 \label{eq:app:evo0}\\
& \begin{aligned}
{x}_2 &= A{x}_1 + B{u}_1 = A(A{{x}}_0 + B{u}_0 ) + B{u}_1 = \\
&= A^2{x}_0 + AB{u}_0 + B{u}_1
\end{aligned}\\
& \begin{aligned}
{x}_3 &= A{x}_2 + B{u}_2 = \\
&= A^3{x}_0 + A^2B{u}_0 + AB{u}_1 + B{u}_2.
\end{aligned}
\end{align}
Iterating until the time interval $N$:
\begin{align}
& \begin{aligned}
{x}_N &= A^N{x}_0 + A^{N-1}B {u}_0 + \cdots + A^0B {u}_{N-1}.
\end{aligned}\label{eq:app:evo2}
\end{align}
Applying \eqref{eq:app:statespace:1} to \eqref{eq:app:evo0}-\eqref{eq:app:evo2} yields:
\begin{align}
&\begin{bmatrix}
 y_0 \\
 y_1 \\
 y_2 \\
 \vdots \\
 y_N
\end{bmatrix}
=
\begin{bmatrix}
 C \\
 CA \\
 CA^2 \\
 \vdots\\
 CA^N
\end{bmatrix}{x}_0 + \\
& + \begin{bmatrix}
 D & 0 & \dots & 0 & 0 \\
 CA^0B & D & \dots & 0 & 0 \\
 CA^1B & CA^0B & \dots & 0 & 0 \\
 \vdots & \vdots & \ddots & \vdots & \vdots \\
 CA^{N-1}B & CA^{N-2}B & \dots & CA^0B & D \\
\end{bmatrix}
\begin{bmatrix}
 u_0 \\
 u_1 \\
 \vdots \\
 u_{N-1}\\
 u_N
\end{bmatrix}\nonumber
\end{align}
which we write in compact form as:
\begin{align}
 \boldsymbol{y} = \phi {x}_0 + \psi_u \boldsymbol{u}.
\end{align}
where $\boldsymbol{y}=y_0, \dots, y_N$, ${x}=x_0, \dots, x_N$ and $\boldsymbol{u}=u_0, \dots, u_N$. For the case of multiple inputs, we add an input $\boldsymbol{r}=r_0, \dots, r_N$ to the state-space model \eqref{eq:app:evo0}-\eqref{eq:app:evo2}:
\begin{align}
 & {x}_{k+1} = A{x}_k + B_u u_k + B_r r_k\\
 & y_k = C{x}_k + D_u u_k + D_u r_k.
\end{align}
The system output is written by applying the transformation $\psi_r$ to $\boldsymbol{r}$:
\begin{align}
 \boldsymbol{y} = \phi {x}_0 + \psi_u \boldsymbol{u} + \psi_r \boldsymbol{r}.
\end{align}

\section*{Acknowledgements}
The authors gratefully acknowledge the support of the colleagues Mr. Marco Pignati and Dr. Paolo Romano for enabling the access to the real-time measurements of the monitored MV feeder of the EPFL campus, and Prof.~Jean-Yves Le Boudec for the useful indications he provided on the integration of BESS constraints in the day-ahead problem.

\bibliographystyle{IEEEtran}
\bibliography{biblio}

\begin{thebibliography}{10}
\providecommand{\url}[1]{#1}
\csname url@samestyle\endcsname
\providecommand{\newblock}{\relax}
\providecommand{\bibinfo}[2]{#2}
\providecommand{\BIBentrySTDinterwordspacing}{\spaceskip=0pt\relax}
\providecommand{\BIBentryALTinterwordstretchfactor}{4}
\providecommand{\BIBentryALTinterwordspacing}{\spaceskip=\fontdimen2\font plus
\BIBentryALTinterwordstretchfactor\fontdimen3\font minus
  \fontdimen4\font\relax}
\providecommand{\BIBforeignlanguage}[2]{{%
\expandafter\ifx\csname l@#1\endcsname\relax
\typeout{** WARNING: IEEEtran.bst: No hyphenation pattern has been}%
\typeout{** loaded for the language `#1'. Using the pattern for}%
\typeout{** the default language instead.}%
\else
\language=\csname l@#1\endcsname
\fi
#2}}
\providecommand{\BIBdecl}{\relax}
\BIBdecl

\bibitem{biegel2014aggregation}
B.~Biegel, P.~Andersen, J.~Stoustrup, M.~B. Madsen, L.~H. Hansen, L.~H.
  Rasmussen \emph{et~al.}, ``Aggregation and control of flexible consumers--a
  real life demonstration,'' in \emph{Proceedings of the 19th IFAC World
  Congress, Cape Town, South Africa}, 2014.

\bibitem{5558756}
J.~Saraiva and M.~Gomes, ``Provision of some ancillary services by microgrid
  agents,'' in \emph{Energy Market (EEM), 2010 7th International Conference on
  the European}, June 2010.

\bibitem{Soshinskaya2014659}
M.~Soshinskaya, W.~H. Crijns-Graus, J.~M. Guerrero, and J.~C. Vasquez,
  ``Microgrids: Experiences, barriers and success factors,'' \emph{Renewable
  and Sustainable Energy Reviews}, vol.~40, 2014.

\bibitem{Bernstein2015}
A.~Bernstein, L.~Reyes-Chamorro, J.-Y.~L. Boudec, and M.~Paolone, ``A
  composable method for real-time control of active distribution networks with
  explicit power setpoints. part i: Framework,'' \emph{Electric Power Systems
  Research}, 2015.

\bibitem{costanzo2013coordination}
G.~T. Costanzo, O.~Gehrke, D.~E.~M. Bondy, F.~Sossan, H.~Bindner, J.~Parvizi,
  and H.~Madsen, ``A coordination scheme for distributed model predictive
  control: Integration of flexible {DERs},'' in \emph{Innovative Smart Grid
  Technologies Europe (ISGT EUROPE), 2013 4th IEEE/PES}.\hskip 1em plus 0.5em
  minus 0.4em\relax IEEE, 2013, pp. 1--5.

\bibitem{7038106}
F.~Ruelens, B.~Claessens, S.~Vandael, S.~Iacovella, P.~Vingerhoets, and
  R.~Belmans, ``Demand response of a heterogeneous cluster of electric water
  heaters using batch reinforcement learning,'' in \emph{Power Systems
  Computation Conference (PSCC), 2014}, Aug 2014.

\bibitem{teleke2009control}
S.~Teleke, M.~E. Baran, A.~Q. Huang, S.~Bhattacharya, and L.~Anderson,
  ``Control strategies for battery energy storage for wind farm dispatching,''
  \emph{Energy Conversion, IEEE Transactions on}, vol.~24, 2009.

\bibitem{marinelli2014testing}
M.~Marinelli, F.~Sossan, G.~T. Costanzo, and H.~W. Bindner, ``Testing of a
  predictive control strategy for balancing renewable sources in a microgrid,''
  \emph{IEEE Transactions on Sustainable Energy}, 2014.

\bibitem{6913566}
M.~Abu~Abdullah, K.~Muttaqi, D.~Sutanto, and A.~Agalgaonkar, ``An effective
  power dispatch control strategy to improve generation schedulability and
  supply reliability of a wind farm using a battery energy storage system,''
  \emph{Sustainable Energy, IEEE Transactions on}, vol.~6, 2015.

\bibitem{5590013}
N.~Troy and M.~O'Malley, ``Multi-mode operation of combined cycle gas turbines
  with increasing wind penetration,'' in \emph{Power and Energy Society General
  Meeting, 2010 IEEE}, 2010.

\bibitem{lu2013nv}
S.~Lu, P.~V. Etingov, D.~Meng, X.~Guo, C.~Jin, and N.~A. Samaan, ``Nv energy
  large-scale photovoltaic integration study: Intra-hour dispatch and agc
  simulation,'' Pacific Northwest National Laboratory (PNNL), Richland, WA
  (US), Tech. Rep., 2013.

\bibitem{EPFL-CONF-203775}
M.~Pignati, M.~Popovic, S.~Barreto~Andrade, R.~Cherkaoui, D.~Flores, J.-Y.
  Le~Boudec, M.~M. Maaz, M.~Paolone, P.~Romano, S.~Sarri, T.~T. Tesfay, D.-C.
  Tomozei, and L.~Zanni, ``Real-{T}ime {S}tate {E}stimation of the
  {EPFL}-{C}ampus {M}edium-{V}oltage {G}rid by {U}sing {PMU}s,'' in \emph{The
  {S}ixth {C}onference on {I}nnovative {S}mart {G}rid {T}echnologies
  ({ISGT}2015)}, 2014.

\bibitem{balancegroup}
``{Balance Group Model (BGM) Introduction},'' SwissGrid, Tech. Rep., as seen on
  June 2016.

\bibitem{6009220}
D.~Wu, D.~Aliprantis, and L.~Ying, ``Load scheduling and dispatch for
  aggregators of plug-in electric vehicles,'' \emph{Smart Grid, IEEE
  Transactions on}, vol.~3, 2012.

\bibitem{6740918}
L.~Yang, J.~Zhang, and H.~Poor, ``Risk-aware day-ahead scheduling and real-time
  dispatch for electric vehicle charging,'' \emph{Smart Grid, IEEE Transactions
  on}, vol.~5, 2014.

\bibitem{7308089}
M.~Bozorg, A.~Ahmadi-Khatir, and R.~Cherkaoui, ``Developing offer curves for an
  electric railway company in reserve markets based on robust energy and
  reserve scheduling,'' \emph{Power Systems, IEEE Transactions on}, 2015.

\bibitem{marinelli2013testing}
M.~Marinelli, F.~Sossan, G.~Costanzo, and H.~Bindner, ``Testing of a predictive
  control strategy for balancing renewable sources in a microgrid,'' \emph{IEEE
  Transactions on Sustainable Energy}, 2014.

\bibitem{teng2013optimal}
J.-H. Teng, S.-W. Luan, D.-J. Lee, and Y.-Q. Huang, ``Optimal
  charging/discharging scheduling of battery storage systems for distribution
  systems interconnected with sizeable pv generation systems,'' \emph{Power
  Systems, IEEE Transactions on}, 2013.

\bibitem{5545425}
S.~Teleke, M.~Baran, S.~Bhattacharya, and A.~Huang, ``Rule-based control of
  battery energy storage for dispatching intermittent renewable sources,''
  \emph{Sustainable Energy, IEEE Transactions on}, vol.~1, 2010.

\bibitem{6417004}
R.~Palma-Behnke, C.~Benavides, F.~Lanas, B.~Severino, L.~Reyes, J.~Llanos, and
  D.~Sáez, ``A microgrid energy management system based on the rolling horizon
  strategy,'' \emph{IEEE Transactions on Smart Grid}, vol.~4, no.~2, June 2013.

\bibitem{bacher2013}
P.~Bacher, H.~Madsen, H.~Nielsen, and B.~Perers, ``Short-term heat load
  forecasting for single family houses,'' \emph{Energy and Buildings}, vol.~65,
  pp. 101--112, 2013.

\bibitem{7232358}
B.~Hayes, J.~Gruber, and M.~Prodanovic, ``Short-term load forecasting at the
  local level using smart meter data,'' in \emph{PowerTech, 2015 IEEE
  Eindhoven}, June 2015, pp. 1--6.

\bibitem{Sossan20161}
F.~Sossan, V.~Lakshmanan, G.~T. Costanzo, M.~Marinelli, P.~J. Douglass, and
  H.~Bindner, ``Grey-box modelling of a household refrigeration unit using time
  series data in application to demand side management,'' \emph{Sustainable
  Energy, Grids and Networks}, vol.~5, 2016.

\bibitem{1350819}
B.-J. Chen, M.-W. Chang, and C.-J. lin, ``Load forecasting using support vector
  machines: a study on eunite competition 2001,'' \emph{IEEE Transactions on
  Power Systems}, vol.~19, no.~4, pp. 1821--1830, Nov 2004.

\bibitem{26540804}
F.~Kasten and G.~Czeplak, ``{Solar and terrestrial radiation dependent on the
  amount and type of cloud},'' \emph{Solar Energy}, vol.~24, pp. 177--189,
  1980.

\bibitem{hofierka2002solar}
J.~Hofierka, M.~Suri \emph{et~al.}, ``The solar radiation model for open source
  gis: implementation and applications,'' in \emph{Proceedings of the Open
  source GIS-GRASS users conference}, 2002, pp. 1--19.

\bibitem{neteler2012grass}
M.~Neteler, M.~Bowman, M.~Landa, and M.~Metz, ``{GRASS GIS: a multi-purpose
  Open Source GIS},'' \emph{Environmental Modelling \& Software}, vol.~31, p.
  124–130, 2012.

\bibitem{5530680}
F.~Oldewurtel, A.~Parisio, C.~Jones, M.~Morari, D.~Gyalistras, M.~Gwerder,
  V.~Stauch, B.~Lehmann, and K.~Wirth, ``Energy efficient building climate
  control using stochastic model predictive control and weather predictions,''
  in \emph{American Control Conference (ACC), 2010}, 2010, pp. 5100--5105.

\bibitem{Hredzak201584}
B.~Hredzak, V.~G. Agelidis, and G.~Demetriades, ``Application of explicit model
  predictive control to a hybrid battery-ultracapacitor power source,''
  \emph{Journal of Power Sources}, vol. 277, pp. 84 -- 94, 2015.

\bibitem{boyd_convexoptimization}
S.~Boyd and L.~Vandenberghe, \emph{Convex Optimization}.\hskip 1em plus 0.5em
  minus 0.4em\relax Cambridge University Press, 2004.

\bibitem{1634598}
M.~Chen and G.~Rincon-Mora, ``Accurate electrical battery model capable of
  predicting runtime and i-v performance,'' \emph{Energy Conversion, IEEE
  Transactions on}, vol.~21, June 2006.

\bibitem{YannLiaw2004835}
B.~Y. Liaw, G.~Nagasubramanian, R.~G. Jungst, and D.~H. Doughty, ``Modeling of
  lithium ion cells—a simple equivalent-circuit model approach,'' \emph{Solid
  State Ionics}, vol. 175, 2004.

\bibitem{bahramipanahenhanced}
M.~Bahramipanah, D.~Torregrossa, R.~Cherkaoui, and M.~Paolone, ``Enhanced
  electrical model of lithium-based batteries accounting the charge
  redistribution effect,'' in \emph{Power Systems Computation Conference
  (PSCC)}, 2014.

\bibitem{4336b714bb4b4ad8869e62d0bf115bf5}
N.~Kristensen, H.~Madsen, and S.~Jørgensen, ``Parameter estimation in
  stochastic grey-box models,'' \emph{Automatica}, vol.~40, no.~2, 2004.

\bibitem{CTSM}
N.~R. Kristensen and H.~Madsen, ``Continuous time stochastic modelling,'' 2003.

\bibitem{1101743}
C.~V. Loan, ``Computing integrals involving the matrix exponential,''
  \emph{IEEE Transactions on Automatic Control}, vol.~23, no.~3, pp. 395--404,
  Jun 1978.

\bibitem{STENGEL}
R.~Stengel, \emph{Optimal Control and Estimation}.\hskip 1em plus 0.5em minus
  0.4em\relax Dover Publications, 1994.

\bibitem{belvedere2012microcontroller}
B.~Belvedere, M.~Bianchi, A.~Borghetti, C.~A. Nucci, M.~Paolone, and
  A.~Peretto, ``A microcontroller-based power management system for standalone
  microgrids with hybrid power supply,'' \emph{Sustainable Energy, IEEE
  Transactions on}, vol.~3, no.~3, pp. 422--431, 2012.

\bibitem{kraning2011operation}
M.~Kraning, Y.~Wang, E.~Akuiyibo, and S.~Boyd, ``Operation and configuration of
  a storage portfolio via convex optimization,'' in \emph{Proceedings of the
  18th IFAC World Congress}, vol.~18.\hskip 1em plus 0.5em minus 0.4em\relax
  Citeseer, 2011, pp. 10\,487--10\,492.

\end{thebibliography}

\begin{IEEEbiographynophoto}{Fabrizio Sossan}
(M'05–SM'07) is an Italian citizen and was born in Genova in 1985. He got his M.Sc. in Computer Engineering from the University of Genova in 2010, and, in 2014, the Ph.D. in Electrical Engineering from the Danish Technical University with the thesis \emph{Indirect control of flexible demand for power system applications}. Since 2015, he is a postdoctoral fellow at EPFL, Switzerland. His main research interest are modelling and optimization applied to power system.
\end{IEEEbiographynophoto}

\begin{IEEEbiographynophoto}{Emil Namor}
is an Italian citizen and was born in Cividale del Friuli in 1989. He got his M.Sc in Electrical Engineering from the University of Padova and from the Ecole Centrale de Lille in 2014.  Since 2015, he is a Ph.D. student at EPFL. His main research interest are modeling and control of battery energy storage systems.
\end{IEEEbiographynophoto}

\begin{IEEEbiographynophoto}{Rachid Cherkaoui}
(M'05–SM'07) received the M.Sc. and Ph.D. degrees from EPFL, Lausanne, Switzerland, in 1983 and 1992, respectively. He is currently the Head of the Power System Group, EPFL, Lausanne, Switzerland. His research interests include electricity market regulation, distributed generation and storage and power system vulnerability mitigation. He is the author of more than 80 scientific publications.
\end{IEEEbiographynophoto}

\begin{IEEEbiographynophoto}{Mario Paolone}
(M’07–SM’10) was born in Italy in 1973. He received the M.Sc. (Hons.) degree in electrical engineering and the Ph.D. degree from the University of Bologna, Bologna, Italy, in 1998 and 2002, respectively. He was a Researcher of Electric Power Systems at the University of Bologna in 2005, where he was with the Power Systems Laboratory until 2011. In 2010, he was an Associate Professor with Politecnico di Milano, Milano, Italy. He is currently an Associate Professor with the Swiss Federal Institute of Technology, Lausanne, Switzerland, where he was the EOS Holding Chair of the Distributed Electrical Systems Laboratory. He has authored and co-authored more than 170 scientific papers published in reviewed journals and presented at international conferences. His current research interests include power systems with particular reference to real-time monitoring and operation, power system protections, power systems dynamics, and power system transients. Dr. Paolone is the Secretary and a member of several IEEE and Cigré Working Groups. He was the Co-Chairperson of the Technical Committee of the 9th edition of the International Conference of Power Systems Transients in 2009. He was a recipient of the IEEE EMC Society Technical Achievement Award in 2013.
\end{IEEEbiographynophoto}
\vfill

\end{document}